\documentclass[pra,twocolumn,superscriptaddress,showpacs,nofootinbib]{revtex4}
\usepackage{graphicx}
\usepackage{amssymb}
\usepackage{epstopdf}
\usepackage{amsmath}
\usepackage{amstext}
\usepackage{latexsym}
\newcommand{\ket}[1]{\left\vert#1\right\rangle}
\newcommand{\bra}[1]{\left\langle#1\right\vert}

\begin{document}

\title{Optimal control of atom transport for quantum gates in optical lattices}
\author{G. De Chiara}
\affiliation{ECT*, BEC-CNR-INFM \& Universit\`a di Trento, 38050
Povo (TN), Italy}
\affiliation{Grup d'Optica, Departament de Fisica, Universitat Autonoma de
Barcelona, 08193 Bellaterra, Spain}
\author{T. Calarco}
\affiliation{ECT*, BEC-CNR-INFM \& Universit\`a di Trento,
38050 Povo (TN), Italy}
\affiliation{Institute for Quantum Information Processing, University of Ulm, D-89069 Ulm, Germany}
\author{M. Anderlini}
\affiliation{INFN, Sezione di Firenze, via Sansone 1, I-50019
Sesto Fiorentino (FI), Italy}
\author{S. Montangero}
\affiliation{NEST-CNR-INFM \& Scuola Normale Superiore, P.zza dei Cavalieri 7
56126 Pisa Italy}
\author{P. J. Lee}
\affiliation{National Institute of Standards and Technology,
Gaithersburg, Maryland 20899, USA}
\author{B. L. Brown}
\affiliation{National Institute of Standards and Technology,
Gaithersburg, Maryland 20899, USA}
\author{W. D. Phillips}
\affiliation{National Institute of Standards and Technology,
Gaithersburg, Maryland 20899, USA}
\author{J. V. Porto}
\affiliation{National Institute of Standards and Technology,
Gaithersburg, Maryland 20899, USA}

\date{\today}

\begin{abstract}
By means of optimal control techniques we model and optimize the
manipulation of the external quantum state (center-of-mass motion) of atoms trapped in
adjustable optical potentials. We consider in detail the cases of
both non interacting and interacting atoms moving between
neighboring sites in a lattice of a double-well optical potentials.  Such a lattice can perform interaction-mediated entanglement of atom pairs and can realize two-qubit quantum gates. The optimized control sequences for the optical potential allow transport faster and with
significantly larger fidelity than is possible with processes based on adiabatic transport.
\end{abstract}


\maketitle

\section{Introduction}
Quantum degenerate gases, such as Bose-Einstein condensates (BECs) \cite{bloch-review} or cold Fermi gases \cite{inguscio-fermi}, trapped in optical lattices, provide a flexible platform for investigating condensed matter models and quantum phase transitions \cite{bloch-qpt}.
It has been proposed to use these systems as
quantum simulators of solid state systems \cite{cirac-chains} and
for implementing quantum information processing (QIP)
\cite{brennen, jaksch, dorner}. Experiments on neutral atoms have
shown some of the ingredients needed for QIP: the preparation of a
Mott insulator state with just one particle per well, which is
used as the initial state of a quantum register \cite{bloch-qpt},
single-qubit rotation \cite{lee07}, and controlled motion of atoms
so as to effect entangling interactions \cite{mandelNature03,lee07}.

A general requirement of QIP is accurate control of a quantum
system. Often this includes control of degrees of freedom other
than the qubit or computational basis, for example the center of mass
motion of an ion or atom for which the spin (internal state) represents the qubit.
One approach to achieving such accurate control is adiabatic
manipulation of the relevant Hamiltonian. Unfortunately
adiabaticity limits the speed of operations. One way to overcome
this difficulty is to use optimal control methods
\cite{oc1,dorner}. Here we show that such techniques could improve
the speed and fidelity of transport of atoms in an optical
lattice.

Recent experiments \cite{anderliniSwap, mandelNature03,trotzky08a}
have shown that quantum gates could be implemented in controllable
optical potentials
by adjusting the overlap between atoms trapped in neighboring
sites of an optical lattice. High fidelity of this  dynamic
process could be achieved by adiabatically changing the trapping
potential. This, however, limits the overall gate speed to be much
lower than the trapping frequency \cite{dorner,garcia-2003}. Here we
present a detailed numerical analysis of the transport process
used to effect a two-qubit quantum gate in \cite{anderliniSwap},
which is performed with the controllable double-well optical
potential described in \cite{anderlini-pra} and find that it gives
an accurate description of the evolution measured in the
experiment.
Then we apply optimal control theory
to the transport process of the atoms, both with and without
interactions, to show how to increase the speed of the gate. The
success of this method in this specific problem demonstrates the
promise of optimal control for coherent manipulation of a diverse
class of quantum systems.

\section{The experiment} \label{sec:model}

A two-qubit quantum gate with neutral atoms can be realized in
optical lattices through a controlled interaction-induced evolution of the wavefunction that depends on the states of the two atoms \cite{brennen,jaksch}.
Because atoms in their electronic ground states generally have short-range interactions, in order to use these contact interactions to produce entanglement, the atomic wavefunctions must be made to overlap.
Once the interaction has taken place for a fixed time, the two atoms
can be separated again thus finishing the gate. In this paper we
consider the control of such motion in a specific setup; however
our theory can be applied to more general systems.

\subsection{The Double-Well Lattice} \label{sec:doublewell}

Neutral $^{87}$Rb atoms are loaded into the sites of a 3D optical
lattice obtained by superimposing a 2D optical lattice of
double-wells \cite{anderlini-pra} in the horizontal plane and an
independent 1D optical lattice in the vertical direction. The
horizontal lattice has a unit cell that can be dynamically
transformed between single-well and double-well configurations.
The horizontal potential experienced by the atoms is
\cite{anderliniJPhysB}:
\begin{eqnarray}\label{eq:potential2D}
V(x,y)&=&- V_0\left[ \cos^2\left(\frac{\beta}{2}\right)(\cos^2k y+\cos^2k
x)+
\right .   \nonumber\\
&+&\left .   \sin^2\left(\frac{\beta}{2}\right)(\cos k y+ \cos(k
x-\theta))^2\right]
\end{eqnarray}
where $x$ and $y$ are the spatial coordinates,
$k=2\pi/\lambda$ is the laser wave-vector and $\lambda$ is the laser wavelength. The potential
\eqref{eq:potential2D} depends on three parameters:
\begin{enumerate}
\item[(i)] the strength $V_0$ of the potential wells; \item[(ii)]
the ratio $\tan\left(\frac{\beta}{2}\right)$ of vertical to horizontal
electric field components;
 \item[(iii)] the phase shift $\theta$ between vertical and horizontal light components.
\end{enumerate}
The angle $\beta$ determines the height of the barrier between
adjacent double-well sites: by changing $\beta/\pi$ from 0 to
$0.5$ the potential changes from a symmetric double-well configuration, with a spacing of lambda/2 (lambda/2 lattice), to a single-well configuration, with a spacing of lambda (lambda lattice). By changing $\beta$ and $\theta$ together
one varies the energy offset (tilt) of a well with respect to the
neighboring one. The tilt of the double well is zero for $\theta
/\pi$ = $\pm 0.5$, while it is maximum (with a value depending on
$\beta$) for $\theta /\pi$ = 0 or $\pm\,1$.

To effect a quantum gate, one varies the three parameters in time so as to move atoms
occupying adjacent wells into the same well, allowing
them to interact and finally returning them to their original
positions.

\begin{figure}[htbp]
  \centering
 \includegraphics*[scale=0.4]{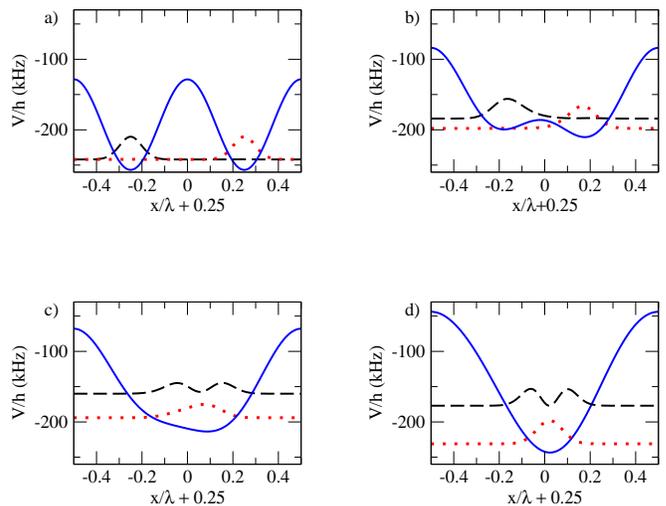}
        \caption{(Color online) a) Initial configuration: the potential (solid lines)
        is in the $\lambda/2$ configuration and the single-particle wave functions
        are localized in the left (dashed) or right (dotted) well.
        b)--c)
        Intermediate snapshots obtained by
        lowering the central barrier and tilting the potential. d) End of the
        process: the particle initially in the right well ends in the ground state of the
        single well, while the particle initially in the left well ends
        in the first excited state. }
      \label{fig:potpsi}
\end{figure}

In Fig.~\ref{fig:potpsi} we show four snapshots of the cross-section
of the optical potential along the direction of the double wells
($x$), and of the single-particle wave functions of the two atoms
during a particular transport sequence. In the initial configuration
each atom is prepared in the ground state of separate wells so that
the properly symmetrized initial state is:
\begin{equation} \label{eq:psiin}
\ket{\Psi_{\rm in}} =\frac{1}{\sqrt 2}\left(
\ket{\psi_L}_1\ket{\psi_R}_2 + \ket{\psi_R}_1\ket{\psi_L}_2\right)
\end{equation}
where $\psi_L$ and $\psi_R$ are wavefunctions localized in the left
and right well, respectively, and 1 and 2 are the labels of the two
(indistinguishable) atoms (see Fig.~\ref{fig:potpsi}a). For quantum gate operation, we would also include the internal state of the atoms, but here we concentrate only on the external state (center-of-mass motion). $\psi_L$ and
$\psi_R$ are linear combinations of the lowest symmetric and
antisymmetric energy eigenfunctions of the single-particle
potential.
In Eq.~\eqref{eq:psiin} and throughout the text we use the convention that single- (two-) particle states are labeled with lower (upper) case greek letters.

The potential is changed by lowering the barrier and at the same
time lowering the right well with respect to the left one, Fig.
\ref{fig:potpsi}b)-c). The atom initially in the right well remains
in the ground state, evolving into the lowest state $\phi_0$ of the
final potential, while the atom initially in the left well evolves
into the first excited state $\phi_1$, Fig. \ref{fig:potpsi}d). When
the two atoms are in the same potential well they interact through
the usual contact interaction, which can
be used to generate the entangling operation needed to realize a two-qubit quantum gate \cite{jaksch,dorner}.


\subsection{Experimental Procedure}
The experimental characterization of the transport process is
accomplished by performing the potential transformation depicted in
Fig.~\ref{fig:potpsi} with atomic samples loaded either only in the
left sites or only in the right sites of the double wells
\cite{anderliniJPhysB}.

Briefly, Bose-Einstein condensates of $^{87}\textrm{Rb}$ atoms
with $4\cdot 10^3 \leq N_{\textrm{BEC}} \leq 2\cdot 10^4$ are
loaded in the sites of the $\lambda$-lattice with an exponential
ramp of 200 ms duration. This loading populates only the ground
band of the optical potential with mostly one atom per lattice
site \cite{SebbyPRL07}. Then the potential is transformed to the
$\lambda/2$-lattice in such a way that the atoms eventually occupy
either only the right sites or only the left sites of the double
wells \cite{anderliniJPhysB,lee07}. Starting from this initialized
state, we perform the transport process illustrated in Fig.
\ref{fig:potpsi}a)--d). At the end of the process we measure the
occupation of the lattice bands. To this purpose, we map the
quasi-momentum of atoms occupying different vibrational levels of
the optical potential onto real momenta lying within different
Brillouin zones \cite{kastberg95,greiner2001}. This is achieved by
switching off the optical potential in 500 $\mu$s and acquiring an
absorption image of the sample after a 13 ms time-of-flight. In
this way atoms occupying different vibrational levels appear
spatially separated, allowing us to measure the amount of
population in each vibrational state.

The comparison between these measurements and the theoretical model
requires an accurate determination of the evolution of the
parameters $V_0$, $\beta$ and $\theta$ characterizing the optical
lattice during the experimental sequences. The parameter $V_0$,
which corresponds to the depth of the potential when it is set in
the $\lambda/2$ configuration, is measured by pulsing the
$\lambda/2$-lattice and observing the resulting momentum
distribution in time of flight \cite{Ovchinnikov1998}. The
parameters $\beta$ and $\theta$, which determine the shape of the
double-well potential and are controlled using two electro-optic
modulators (EOMs), are determined from measurements of the
polarization of the laser beams after the EOMs as a function of
their respective control voltages \cite{anderlini-pra}.

\begin{figure}[htbp]
 \centering
 \includegraphics*[scale=0.5]{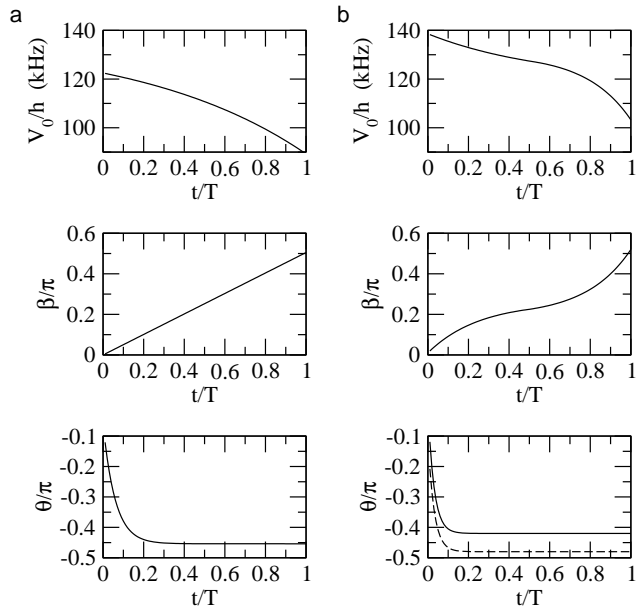}
        \caption{Two possible sequences {\sf a}
        (left) and {\sf b} (right)
        employed to shift the atoms from a double- to a single-well configuration
        are shown in the left and right part of the panel. For each sequence we show
        the time-dependence of $V_0$, $\beta$, $\theta$ for a sequence duration $T$. For sequence {\sf b} we show the time dependence of $\theta$ for two settings of EOM$\theta$: $-0.42\; \pi$ (solid) and $-0.48\; \pi$ (dashed).}
        \label{fig:pulse}
\end{figure}

We perform two series of experimental sequences in order to study
the properties of the atomic transport as a function of
the duration of the process and of the energy tilt between the two
potential wells during the merge. In a first series of
measurements the sequence involves converting
the lattice from the double-well to the single-well
configuration by changing $\beta$, rotating the polarization of
the input light using a linear ramp, while leaving constant the
light intensity and the setting of the electro-optic modulator
EOM$\theta$ dedicated to the control of the phase shift $\theta$.
This sequence is repeated for different durations of the linear
ramp from $T = 0.01$ ms to 1.01 ms. In a second series of
measurements we consider the dependence of the transport on the
tilt of the double-well potential during the merge. We perform the
lattice transformation using always the same duration of $T = 0.5$
ms, the same intensity of the light beam and the same ramp for
changing the polarization angle $\beta$, while the the setting of EOM$\theta$ is kept constant in
time during a sequence. We then repeat the sequence for different
settings of EOM$\theta$. The time dependence for
the three lattice parameters $V_0$, $\beta$ and $\theta$ for
measurements of the first series and of the second series are
shown in Fig. \ref{fig:pulse}{\sf a} and Fig. \ref{fig:pulse}{\sf
b}, respectively.  Fig. \ref{fig:pulse}{\sf b} shows the evolution
of the parameter $\theta$ for two different settings of
EOM$\theta$. The potential parameters are determined using
our calibration of the lattice setup, taking into account effects
such as different losses on the optical elements for different
polarizations of the lattice beams and the dependence of the
optical potential on the local polarization of the light
\cite{lee07}. These effects are responsible for the change of the
potential depth $V_0$ and of the angle $\theta$ during the
sequence despite the fact that both the intensity of the
light and the settings of EOM$\theta$ are not actively changed.

\section{Theoretical model} \label{sec:nint}
Here we describe the theoretical methods that we implement for
investigating the dynamics in the system described above, starting
with the case of non-interacting particles. Then, we
consider the experimental realization of the merging of adjacent
lattice sites into a single site shown in Fig. \ref{fig:potpsi}
and we compare the results obtained in the experiment with the
expectations based on our theoretical model. This stage represents a useful
benchmark to evaluate the reliability of the numerical model as
well as for gaining insight into the details of the optical
potential experienced by the atoms. Finally, we present the technique for optimizing the transport sequence, and we
show how we can achieve a significantly higher fidelity at fixed operation time for the atomic motion than by using smooth sequences based on adiabatic
evolution.

\subsection{Theoretical Framework}
We consider the 1D problem restricted to the x axis by
assuming that the optical potential can be separated along the
three spatial directions, allowing us to express the atomic
wavefunctions as a product of three independent terms.
We consider the harmonic approximation of the
potential in the $y$
and $z$ directions, having trap frequencies $\nu_y$ and $\nu_z$
respectively, that can be calculated as shown in
\cite{spielman06} and we assume that along $y$ and $z$ the
atoms always occupy the lowest vibrational state. This restriction
does not put limitations in studying dynamic processes involving
low energy states of the double-well potential since it can be
chosen to have non-degenerate vibration frequencies along all
three directions, with the lowest frequency always along $x$. We calculate the
eigenstates of the system along the $x$ direction by solving the
eigenvalue equation using the finite difference method
\cite{thomas}. For the time evolution we consider the integration
of the time dependent Schr\"{o}dinger equation using the
Crank-Nicolson method \cite{cn}. This method has the advantage of being unconditionally stable and the error in the results scale quadratically with the number of space-time grid points in which the Schr\"{o}dinger equation is solved. The relative error of the data presented is always less than $10^{-3}$. In Appendix \ref{sec:numerical}
we present a more detailed description of our numerical methods.

\subsection{Comparison to experimental results}
\begin{figure}[htbp]
  \includegraphics*[scale=0.5]{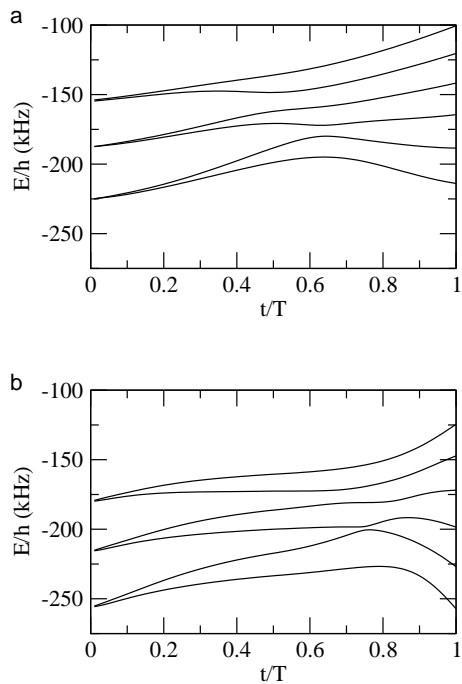}
        \caption{Instantaneous spectrum of the 1D Hamiltonian for
        sequences {\sf a} and {\sf b} for EOM$\theta$: $-0.42\; \pi$.}
      \label{fig:spectrum}
\end{figure}

In this section we present the theoretical analysis of the
transport processes described above and we discuss the agreement
between the model and the experimental measurements. We start by
considering the time evolution of the Hamiltonian spectrum during
the two sequences {\sf a} and {\sf b} shown in Fig.
\ref{fig:spectrum}.

At time $t=0$ the spectrum is made of nearly degenerate doublets of
almost equally spaced pairs of harmonic oscillator states, while at time $t=T$ the levels are similar
to those of a single harmonic oscillator\footnote{We have verified that
the results for the 1D spectrum are in good agreement with full
calculations in 2D (restricted to states with vibrational
excitation along the $x$ direction). Additional energy levels are
present in the 2D spectrum, associated with states with
vibrational excitation along $y$. However, those states can be
neglected for studying the dynamical process considered here since
their energy is always higher than the three lowest states of the
1D spectrum.}.

Fig.~\ref{fig:spectrum} shows the time evolution of the eigenstates of the single-particle Hamiltonian.
The atoms initially prepared in the two local ground states
in the right and left wells ($\psi_R$ and $\psi_L$) evolve into the
instantaneous eigenstates ending in the ground and first excited
state of the final configuration, respectively. This approach
requires the sequence to be performed slowly with respect to the
timescale associated with the gaps between the relevant energy
levels. The optimal ``speed'' in the parameter space can be
calculated using the Landau-Zener theory for avoided level
crossings.

For gaining quantitative insight into the properties of the
transport we perform numerical simulations for the sequences used
in the experiments, also taking into account possible deviations
of the parameters from the experimental calibrations, and we
compare the results with the experimental measurements.
The relevant quantities for our analysis will be the
overlap $f^\alpha_n$ of the energy eigenstates $\phi_{n}$ of the
final potential with the evolved state $\psi_\alpha$ where $\alpha=L,R$ indicates the initial well occupation:
\begin{equation}
f_n^\alpha=\left|\bra{\phi_n} U(T)\ket{ \psi_{\alpha}} \right |^2
\end{equation}
where the operator $U(T)$ is the single-particle time-evolution operator from time $t=0$ to time $t=T$.
In the experiment $f^\alpha_n$ can be measured as
the population of each energy level at the end of the process.

We now consider how the atoms evolve when the parameters
change according to sequence {\sf a} of Fig.~\ref{fig:pulse} as a
function of the total time $T$, focusing on atoms starting in
$\ket{\psi_L}$\footnote{Both in the experiments and in the
simulations the evolution of the atom initially in the right well,
i.e. in state $\ket{\psi_R}$, shows a weaker dependence on the
properties of the sequence and is less instructive. For instance, in
the simulations for $T=0.5$ ms the population in the ground state
$f_0^R$ is of order of $99 \%$ for a broad range of parameters.}.
\begin{figure}[htbp]
\hspace*{2.2in}\mbox{}\\\includegraphics*[scale=0.5]{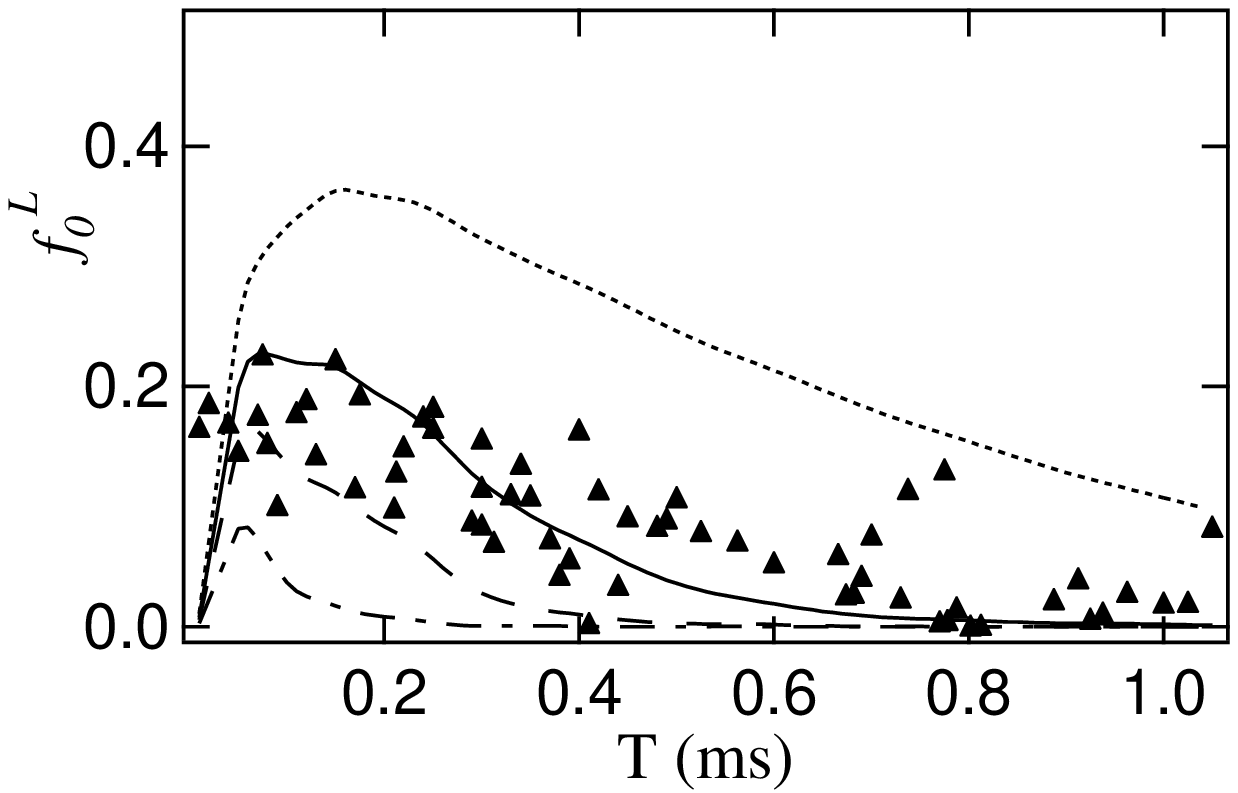}\\
\hspace*{2.2in}\mbox{}\\\includegraphics*[scale=0.5]{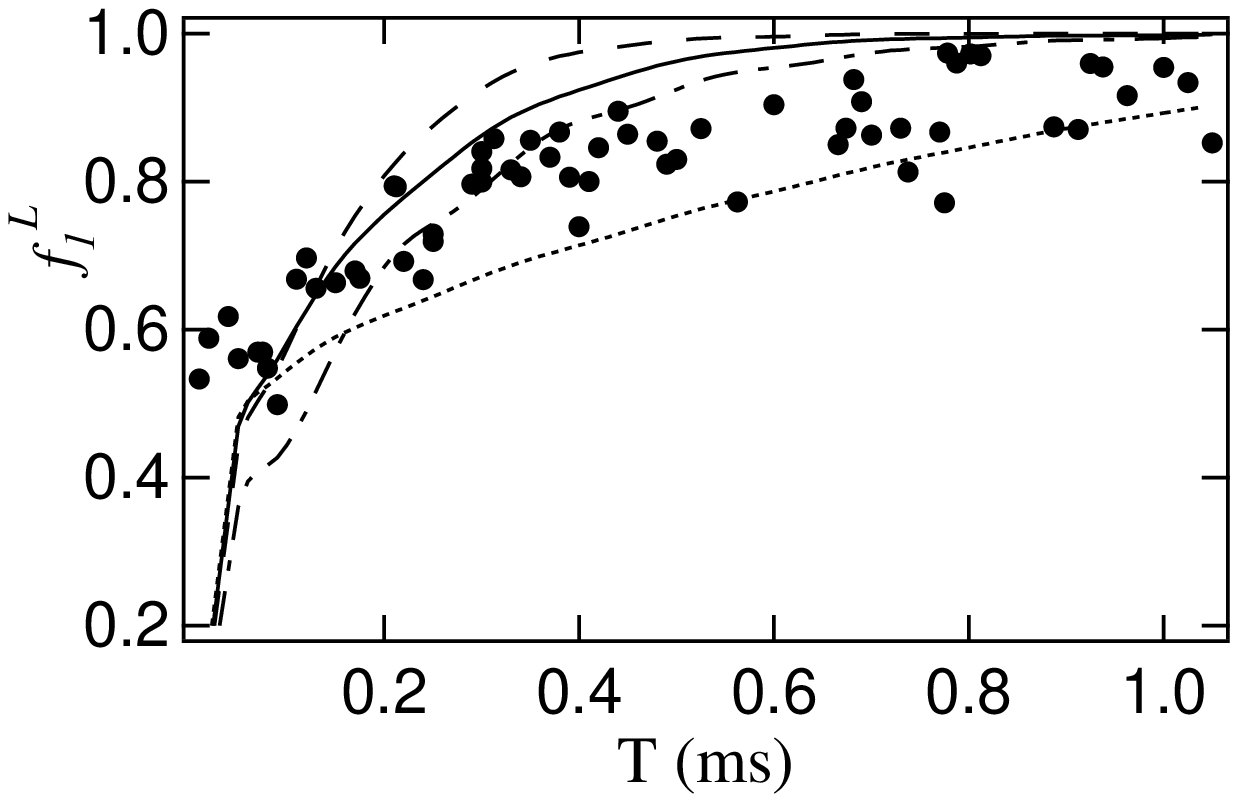}\\
 \hspace*{2.2in}\mbox{}\\\includegraphics*[scale=0.5]{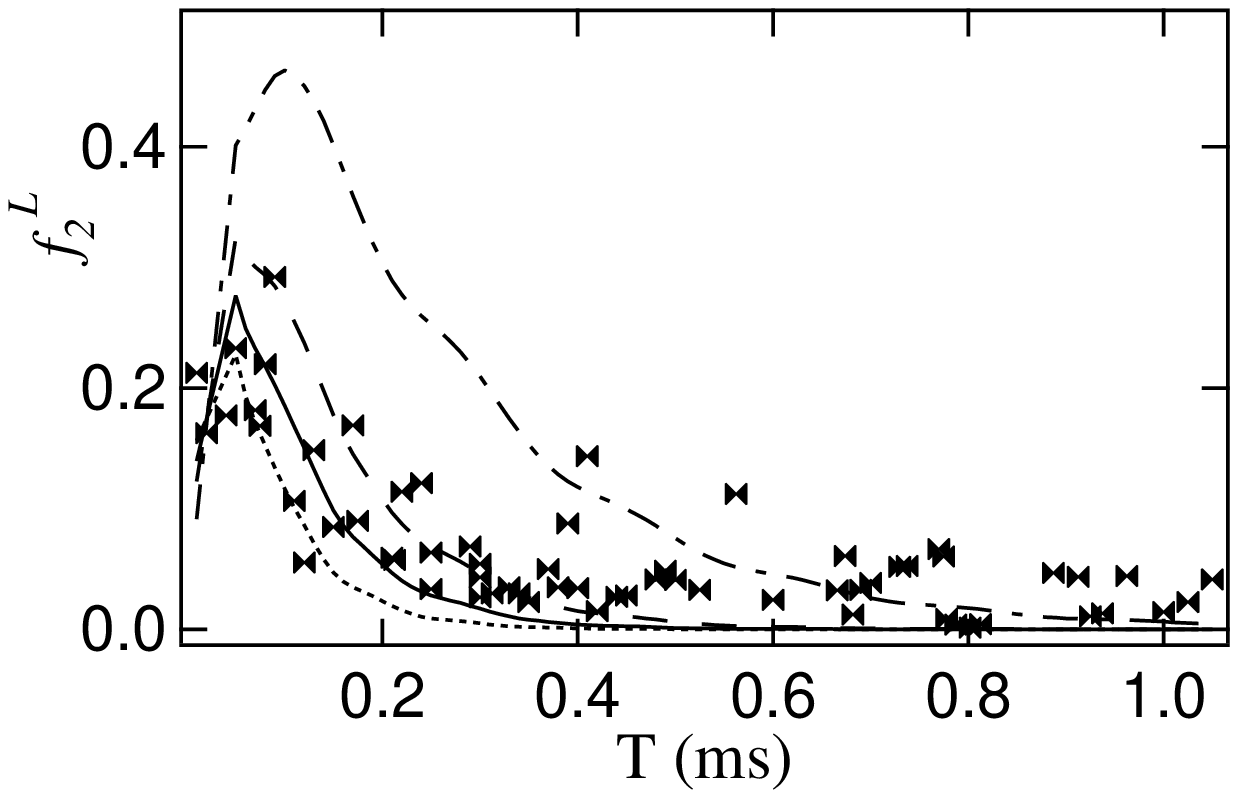}
        \caption{Population  of the first three
        eigenstates of the Hamiltonian, ground (top),
        first excited (middle), second excited (bottom), at the end of sequence {\sf a} as
        a function of the sequence duration $T$.
        The experimental data (symbols) are compared to the four sets of numerical data
        (lines) obtained for $\theta/\pi = -0.454 +\Delta
        \theta_a/\pi$, with $\Delta
        \theta_a/\pi$ = 0 (dot-dash), -0.02 (dash), -0.03 (solid) and
        -0.04 (dot), while -0.454 is the nominal value
        of $\theta/\pi$ expected from the calibrations.
    }
      \label{fig:population1}
\end{figure}
In Fig.~\ref{fig:population1} we show the final population of the
ground $f_0^L$, first $f_1^L$ and second excited state $f_2^L$ measured in
the experiments and calculated for four values of $\theta$ which
differ from the one of Fig.~\ref{fig:pulse} {\sf a} by a constant
offset $\Delta \theta_a$ \footnote{We do not consider variations in
$V_0$ and $\beta$ due to the small dependence of the transport
process on those parameters within the range associated with the
accuracy of our calibrations.}. The results obtained by the model are
in reasonable agreement with the experimental observations; we find best
quantitative matching for $\Delta \theta_a/\pi$ = -0.03, for which
the rms deviation between model and theory is reduced from 0.13(at $\Delta \theta =0$) to
0.08.

\begin{figure}[t]

  \centering
 \includegraphics*[scale=0.45]{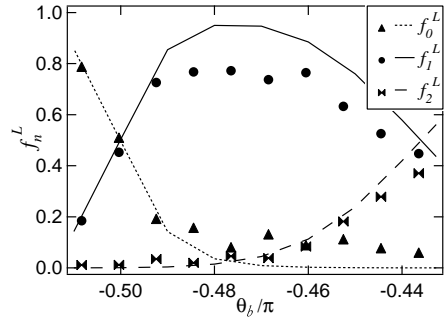}
         \caption{
         Population (overlap absolute squared) of the first three
         eigenstates of the Hamiltonian at the end of sequence {\sf b}
         as a function of $\theta_b$.
         The duration of the sequence is fixed
         to $T = 0.5$ ms. The experimental data (symbols) are in good agreement with
         the numerical data (lines). In this graph the $x$ axis for the experimental
         has been shifted by an offset of -0.015 with respect to the initial
         calibration.
        }
      \label{fig:population2}
\end{figure}

Now we consider the sequence {\sf b} of Fig.
\ref{fig:pulse}, performed for different ending values $\theta_b$ of
the parameter $\theta$ around $-0.47 \pi$. As shown in Fig.~\ref{fig:population2}, both the experiment and
the model show a strong dependence on $\theta_b$ for the transport
of the atom starting in the left site of the double well. The
transport into the first excited state has a maximum theoretical
fidelity of 0.95 for $\theta_b/\pi$ = -0.474. Less negative values
of $\theta_b$, i.e. increasing tilts, lead to a decrease of fidelity
due to the increase in the fraction of population ending in the
second excited state. Values of $\theta_b/\pi$ closer to -0.5, i.e.
more symmetric configurations of the double well, lead to decrease
of fidelity associated with larger fractions of population ending in
the ground state. Also in this case the experimental data and the
theoretical model are in satisfactory agreement. For these data the
deviation between theory and experiment is more sensitive to the
value of the phase shift $\theta_b$. We find best agreement by
shifting the value determined from the calibration by an offset
$\Delta \theta_b/\pi$ = -0.015, which reduces the rms deviation from
0.4 to 0.15. The axis for the experimental data in Fig.~\ref{fig:population2} has been corrected by the offset $\Delta \theta_b$.
Thus, while showing the reliability of the model in describing the
dynamic process, the comparison between theoretical and experimental
results also allows one to refine the calibration of the parameters
characterizing the optical potential.

Finally we find that adding an offset of $\Delta \theta/\pi$ =
-0.016 to the $\theta$ calibration brings the data from both
sequences to a good agreement with the theory and reduces the rms
deviation from 0.19 to 0.11. This is three times larger than the uncertainty of the offset from our EOM calibration but is still consistent with measurements of the lattice tilt from \cite{SebbyPRL07}.  This discrepancy might be explained by the birefringence in the vacuum cell windows, which is not accounted for in our model. Inclusion of this offset should improve both the predictivity
of the model and the experimental optimization of the collisional
gate based on the numerical technique described below.

\section{Optimized transport}

In this section we employ optimal control theory to obtain fast and
high-fidelity gates. Our aim is to find a temporal dependence of the
control parameters $V_0(t)$, $\beta(t)$, $\theta(t)$ that improves
the fidelity even for a shorter sequence duration, when the
adiabatic sequences presented above yield a poor fidelity. Quantum
optimal control techniques have been successfully employed in a
variety of fields: molecular dynamics \cite{oc1, oc2, oc3},
dynamics of ultracold atoms in optical lattices \cite{tannor-2002,vaucher,romero}, implementation of
quantum gates \cite{dorner,monta-oc}.

We use the Krotov algorithm \cite{krotov} as the optimization
procedure. The objective is to find the optimal shapes of the control
parameter sequences that maximize the overlap (fidelity) between the
evolved initial wave function and a target wave function. The
initial and target wave functions are fixed a priori.
The algorithm works also for more than
one particle. The method consists in iteratively modifying the shape
of the control parameters according to a ``steepest descent method''
in the space of functions (for more details see \cite{dorner}). The
method requires evolving each particle's wave function and an
auxiliary wave function backward and forward in time according to the
Schr\"odinger equations. In our simulations we use the
Crank-Nicolson scheme to realize this step as described in
Appendix~\ref{sec:numerical}.

\subsection{Non-interacting case}

\begin{figure}[htbp]
  \centering
 \includegraphics*[scale=0.3]{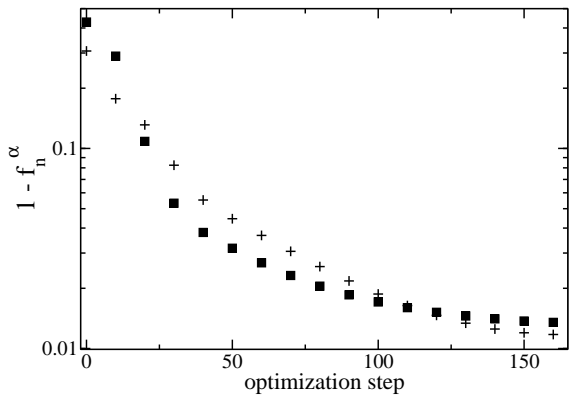}
         \caption{Infidelities ($1-f^\alpha_n$) for the atom initially in the left
         ($\alpha=L, n=1$, squares) and in the right well ($\alpha=R, n=0$, plus) as a function of the optimization step.
           }
      \label{fig:infid-opt}
\end{figure}

We optimize the gate for $T=0.15$ ms\footnote{We chose this time in
order to show the benefits of the optimization procedure for a
sequence duration which cannot provide a good fidelity with smooth
parameter ramps based on adiabatic evolution. While a shorter
duration could be chosen in principle, this choice can be easily
experimentally implemented with no major changes in the present
experimental apparatus, allowing future prosecution of our studies
on this subject.  Increasing the total time should improve the best fidelity.} choosing as a starting point for the optimization a sequence similar to Fig.~\ref{fig:pulse}{\sf b} where $\theta$ is for simplicity taken constant to the final value $\theta_b/\pi=-0.474$. Without optimization the fidelities for the atom
initially in the left and right well are $f^L_1 = 0.57$ and $f_0^R =
0.69$, respectively. The infidelities are shown in
Fig.~\ref{fig:infid-opt} as a function of the number of optimization
steps: the algorithm of optimization is proven to yield a monotonic
increase in fidelity \cite{oc1}, however it does not guarantee to
reach its 100\% value. The results for the two atoms give a fidelity
above $98.7 \%$.

\begin{figure}
  \centering
 \includegraphics*[scale=0.5]{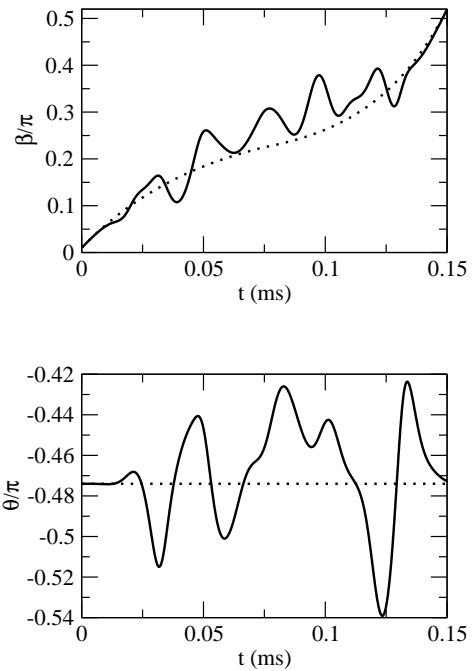}
         \caption{Initial (dotted) and optimized waveforms (solid)
         $\beta(t)$ and $\theta(t)$ as a function of time for a sequence
         of $T=0.15$ ms.}
      \label{fig:pulse-opt}
\end{figure}

The resulting optimized parameter sequences are shown in
Fig.~\ref{fig:pulse-opt} and compared to the original sequence without
optimization.
We find that the optimized sequence for the potential depth $V_0$
differs negligibly from the initial guess.  In principle, the
algorithm could achieve still higher single-particle fidelities from
different starting points.

\begin{figure*}[t!]
  \hspace{-1cm}
  \hspace{-1cm}
 \includegraphics*[scale=0.17,angle=270]{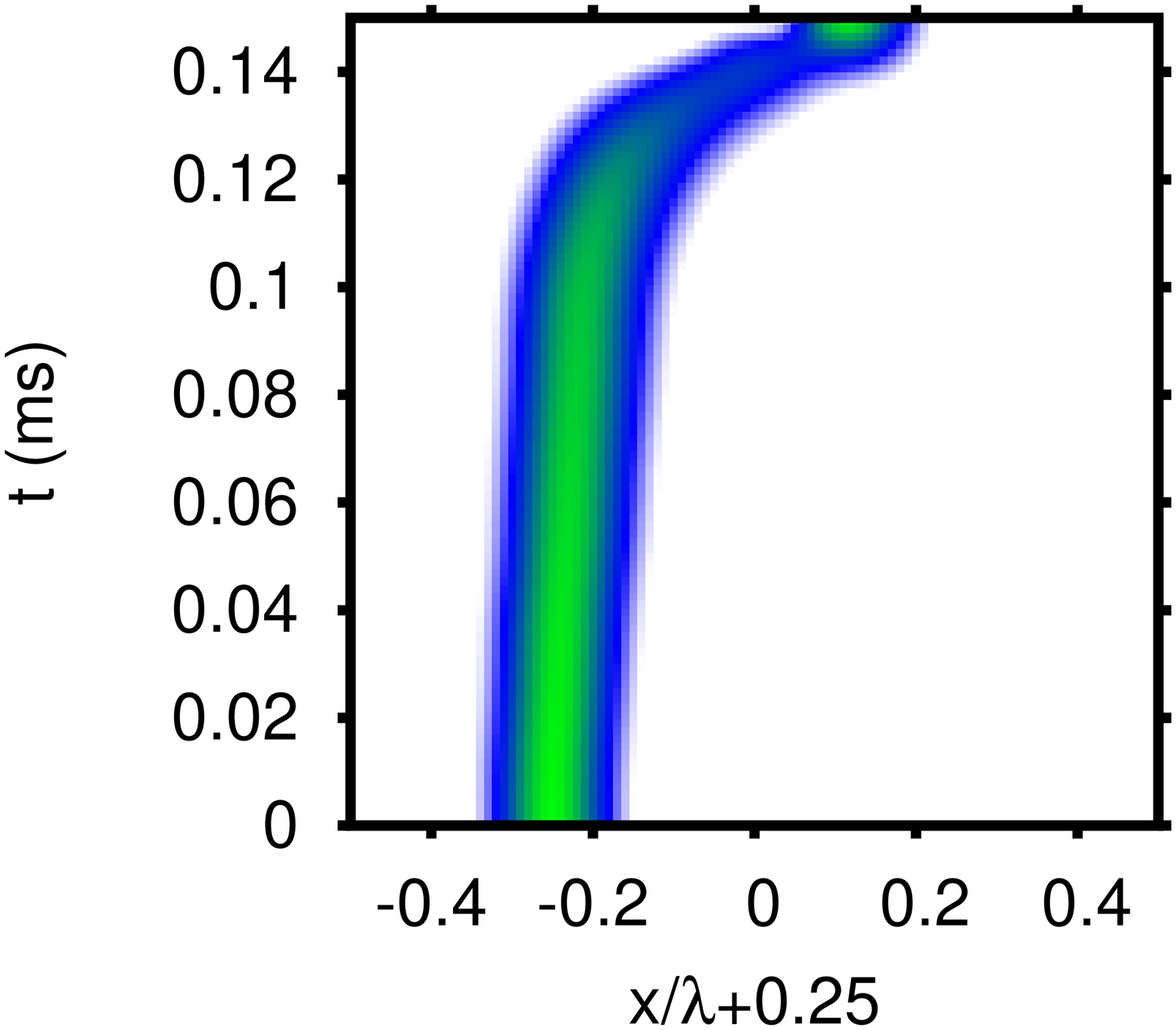}
 \hspace{-1cm}
  \includegraphics*[scale=0.17,angle=270]{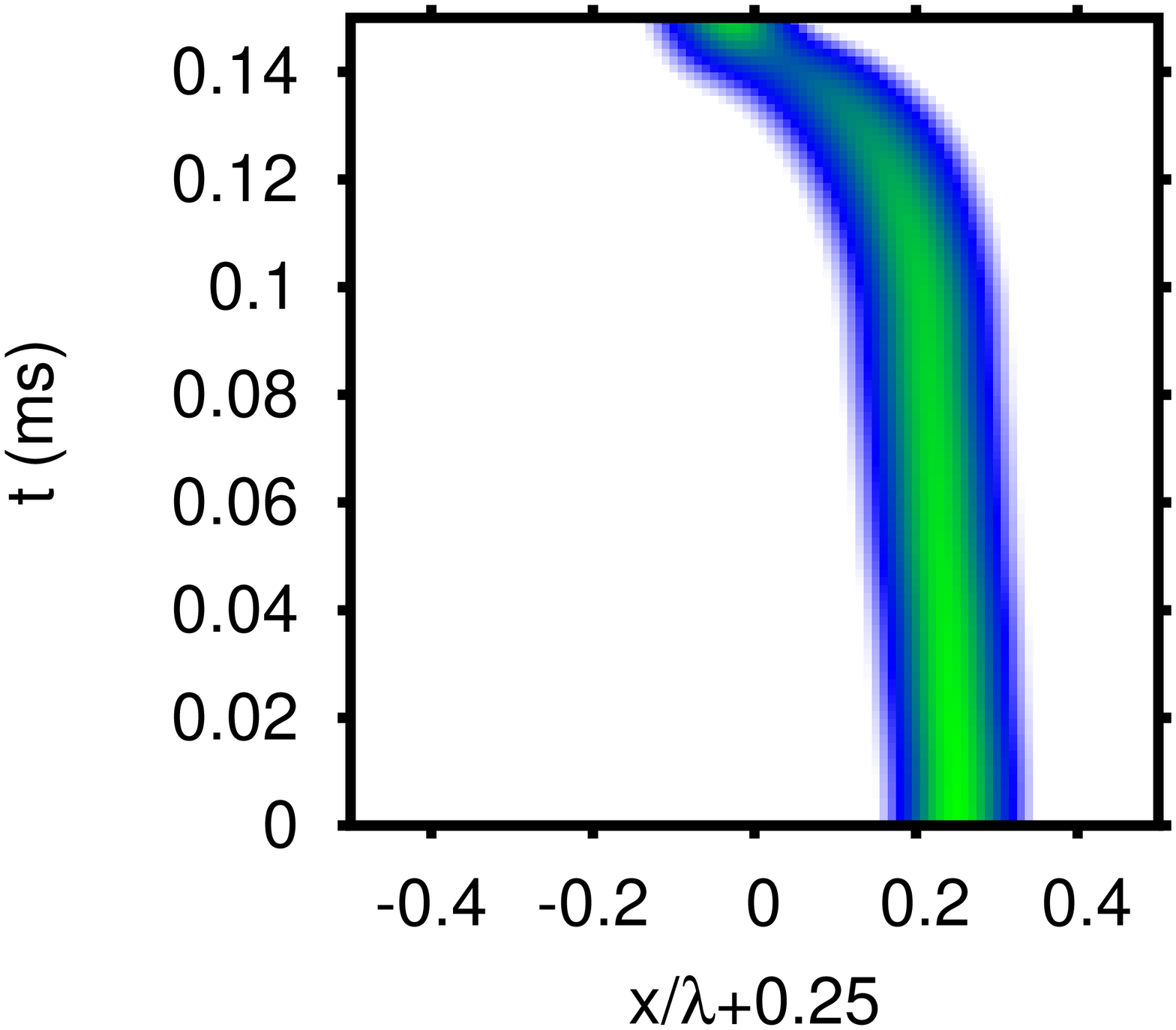}
 \hspace{-1cm}
 \includegraphics*[scale=0.17,angle=270]{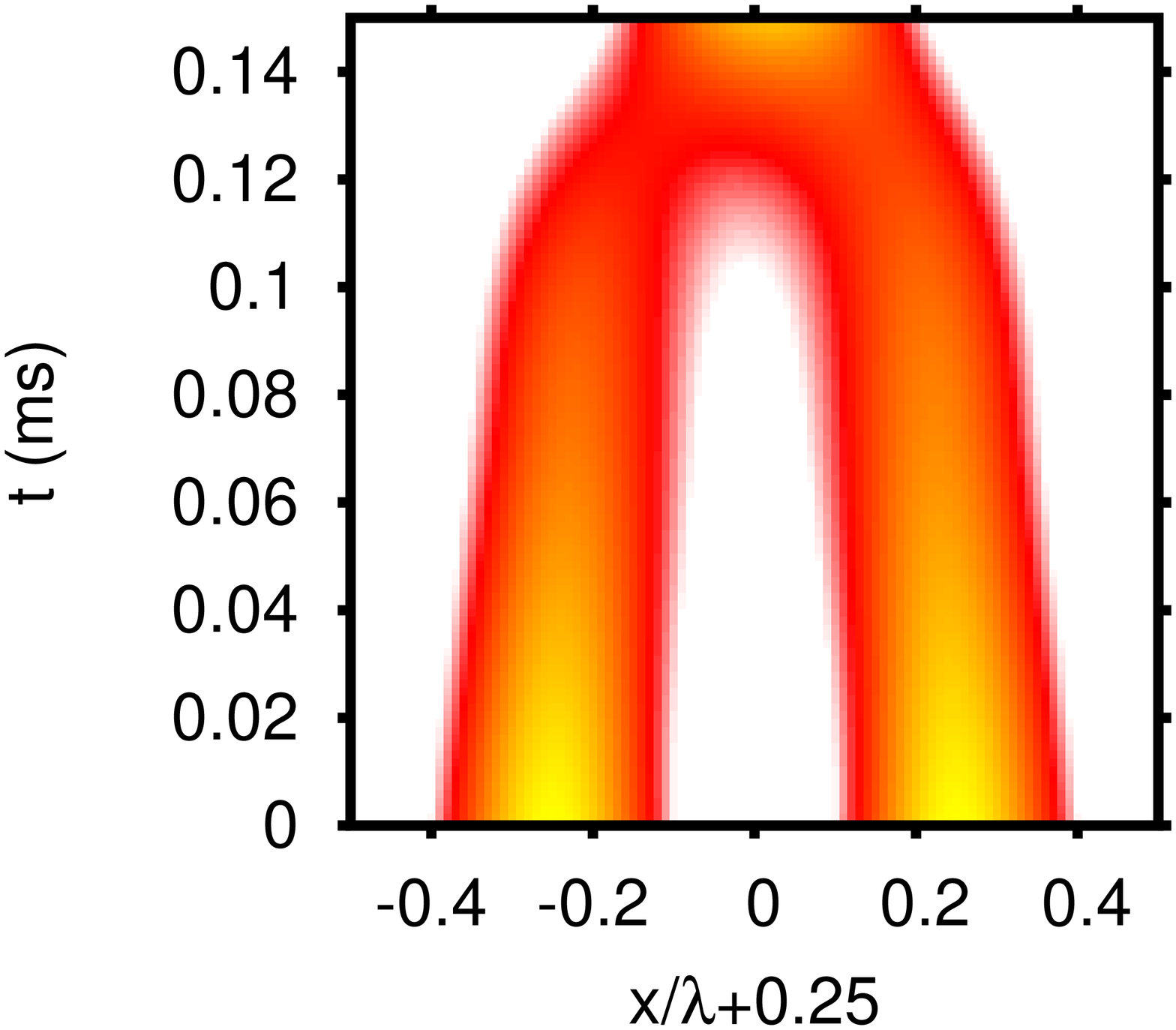}
\hspace{-0.5cm}
 \includegraphics*[scale=0.28,angle=270, viewport=45 30 460 520]{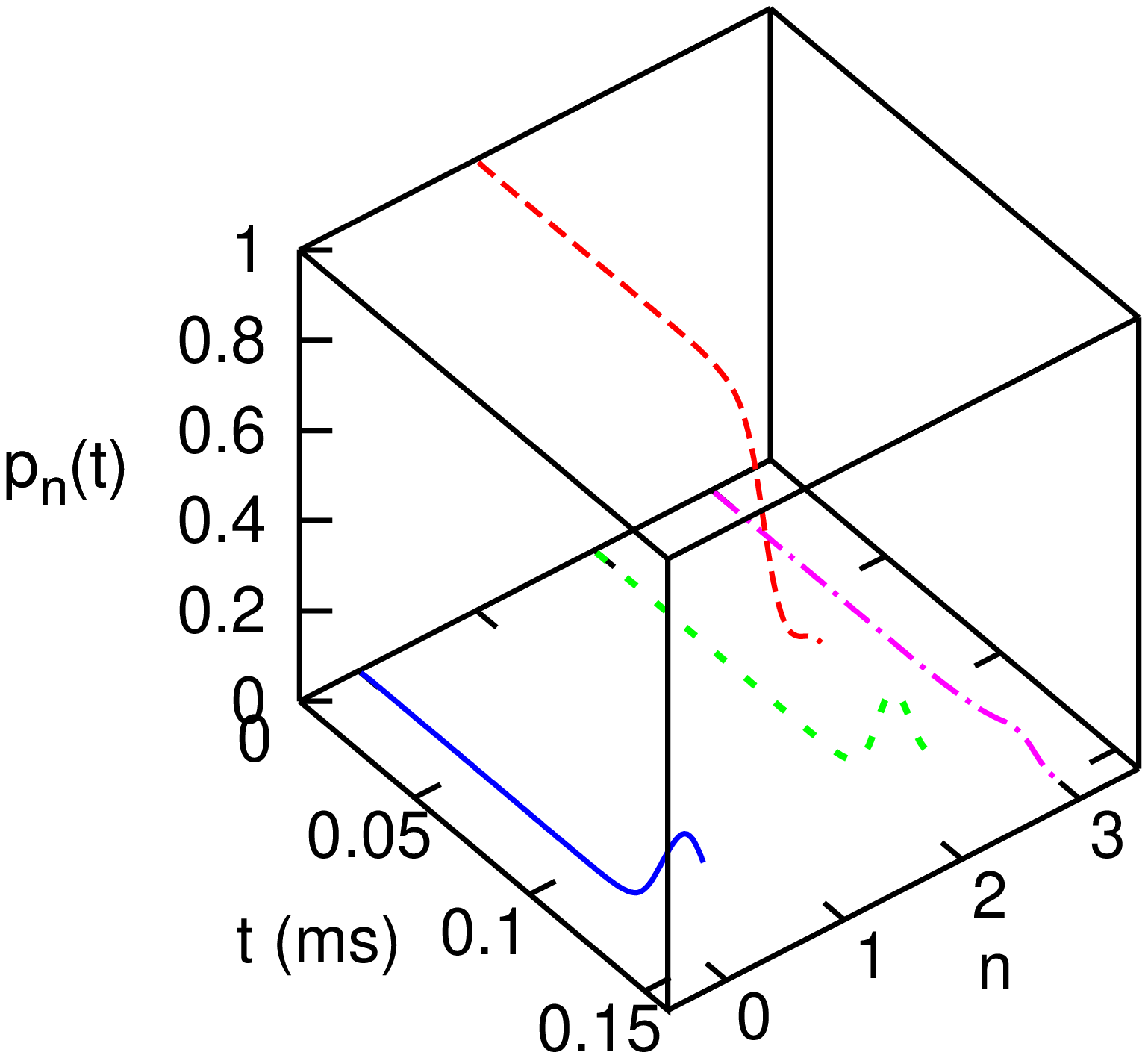}\\
 \vspace{1 cm}
  \hspace{-1cm}
   \hspace{-1cm}
 \includegraphics*[scale=0.17,angle=270]{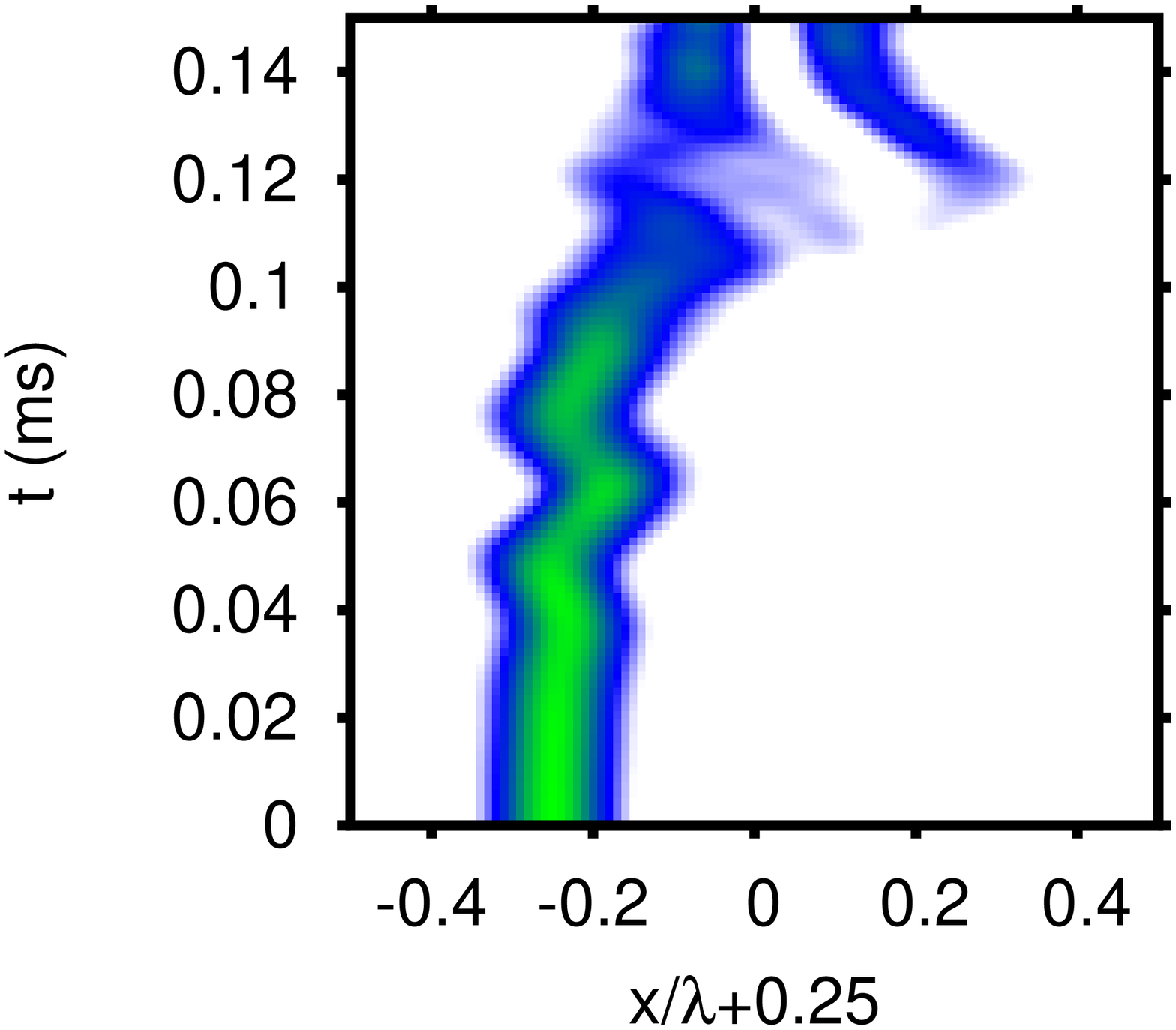}
 \hspace{-1cm}
  \includegraphics*[scale=0.17,angle=270]{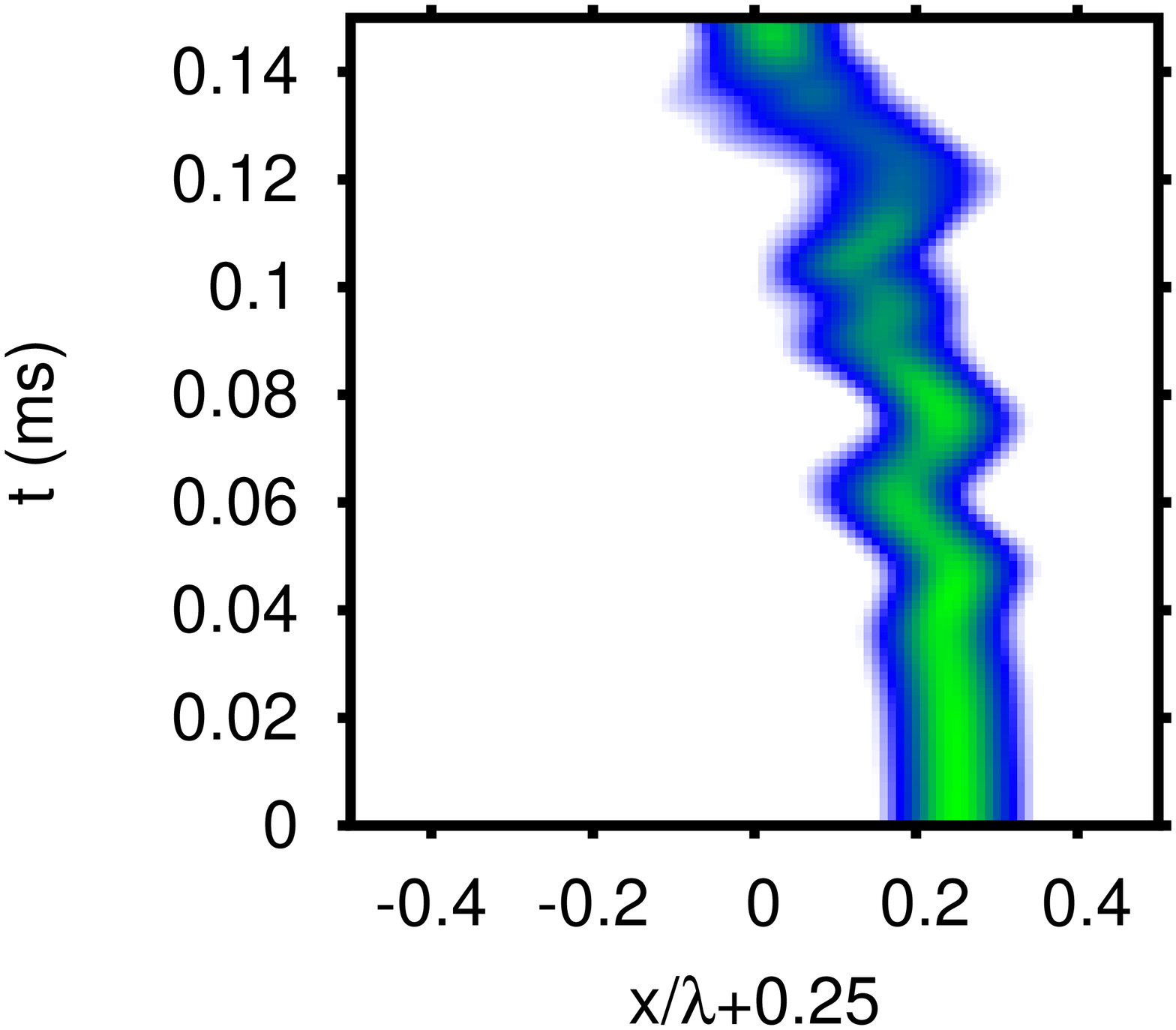}
 \hspace{-1cm}
  \includegraphics*[scale=0.17,angle=270]{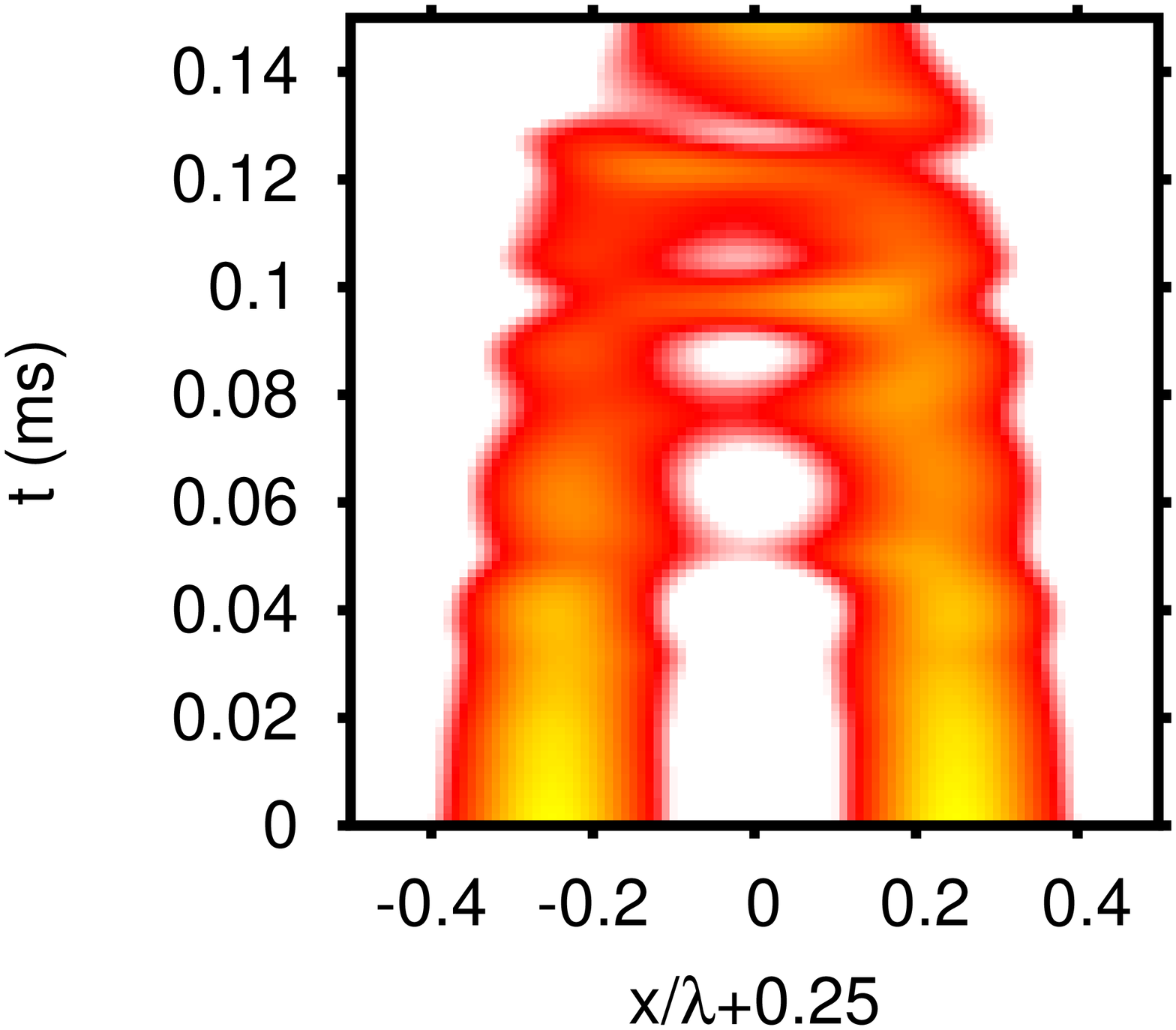}
 \hspace{-0.5cm}
 \includegraphics*[scale=0.28,angle=270, viewport=45 30 460 520]{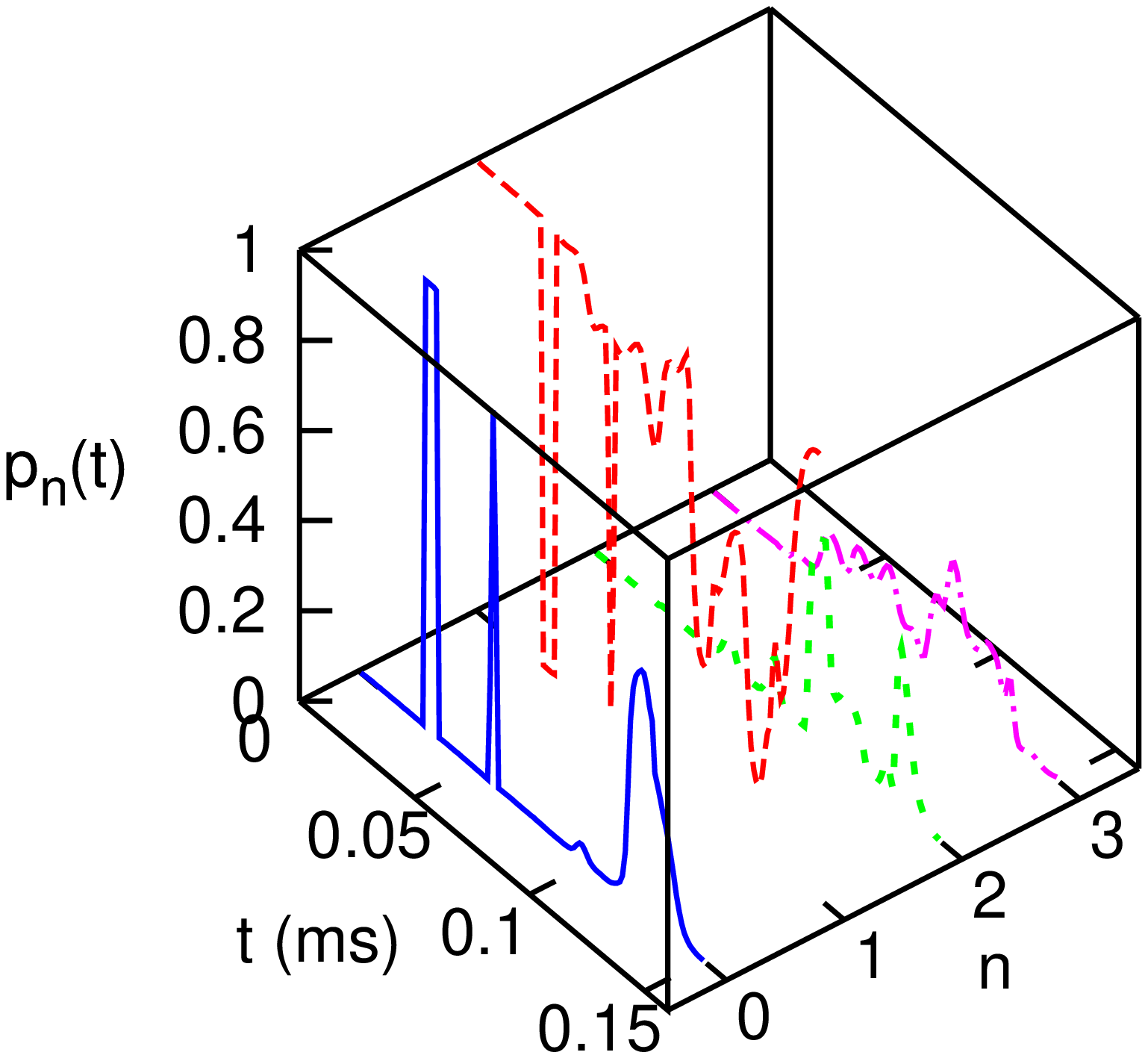}
         \caption{(Color online) Comparison between the evolution of the atoms with and
         without optimal control.  Top (left to right): non optimized case, absolute square value of the wave functions
         as a function of time (atoms initially in the left and right well respectively);  1D trapping potential
         as a function of time; projections $p_n(t)$ at time $t$ of the state initially in the left well onto the instantaneous eigenstates $\ket{\phi_n(t)}$ with $n=0$ (blue solid), $n=1$ (red dashed), $n=2$ (green dotted), $n=3$ (magenta dot-dashed).  Bottom: analogous plots for the optimized case.}
      \label{fig:psi-evol}
\end{figure*}

  In Fig.~\ref{fig:psi-evol} we show the square absolute value of the wave functions of
the two atoms as a function of time, the 1D potential time
dependence and the projections of the initially left-well state onto the lowest four instantaneous energy eigenstates $\ket{\phi_n(t)}$:
\begin{eqnarray}
p_n(t) = \left | \bra{\phi_n(t)} U(t)\ket{\psi_L} \right|^2.
\end{eqnarray}
Notice that $p_n(T)=f_n^L$.  As can be easily seen,
the optimal time evolution is much less smooth than the adiabatic one
as it takes advantage of quantum interference between non-adiabatic excitation paths
to obtain better results.

\subsection{Interaction effects}\label{sec:interactions}
Up to now we have considered only the
single-particle evolution of the system, i.e. without including
any interaction between the particles. This approximation is valid
in our transport sequence as long as the two wave functions
in nearby wells do not overlap. When the two
particles overlap in the same well we must take into account
interactions, and we model them with an effective 1D contact
potential:
\begin{equation}
V_{\rm{int}}(|x_1-x_2|) = g_{1D}\delta(x_1-x_2)
\end{equation}
where $x_i$ are the coordinates of the two atoms and $g_{1D}$ is an
effective $1D$ coupling strength \cite{olshanii} expressed by
$g_{1D} = 2 a_s h \sqrt{\nu_y \nu_z}$, where $a_s$ is the scattering
length for $^{87}\mathrm{Rb}$ atoms and $h$ is the Planck constant.
The spectrum is modified by the interactions: the state with one atom in each well is lower by $\sim 3$ kHz than the doubly-occupied states.

As in the case without interactions we start with each atom
localized in a separate well.
Notice that we are considering wave-functions that are symmetric
under the exchange of the coordinates of the two particles.
\begin{figure}[htbp]
  \centering
 {\sf a}\hspace{-4mm}\includegraphics*[scale=0.17,angle=270]{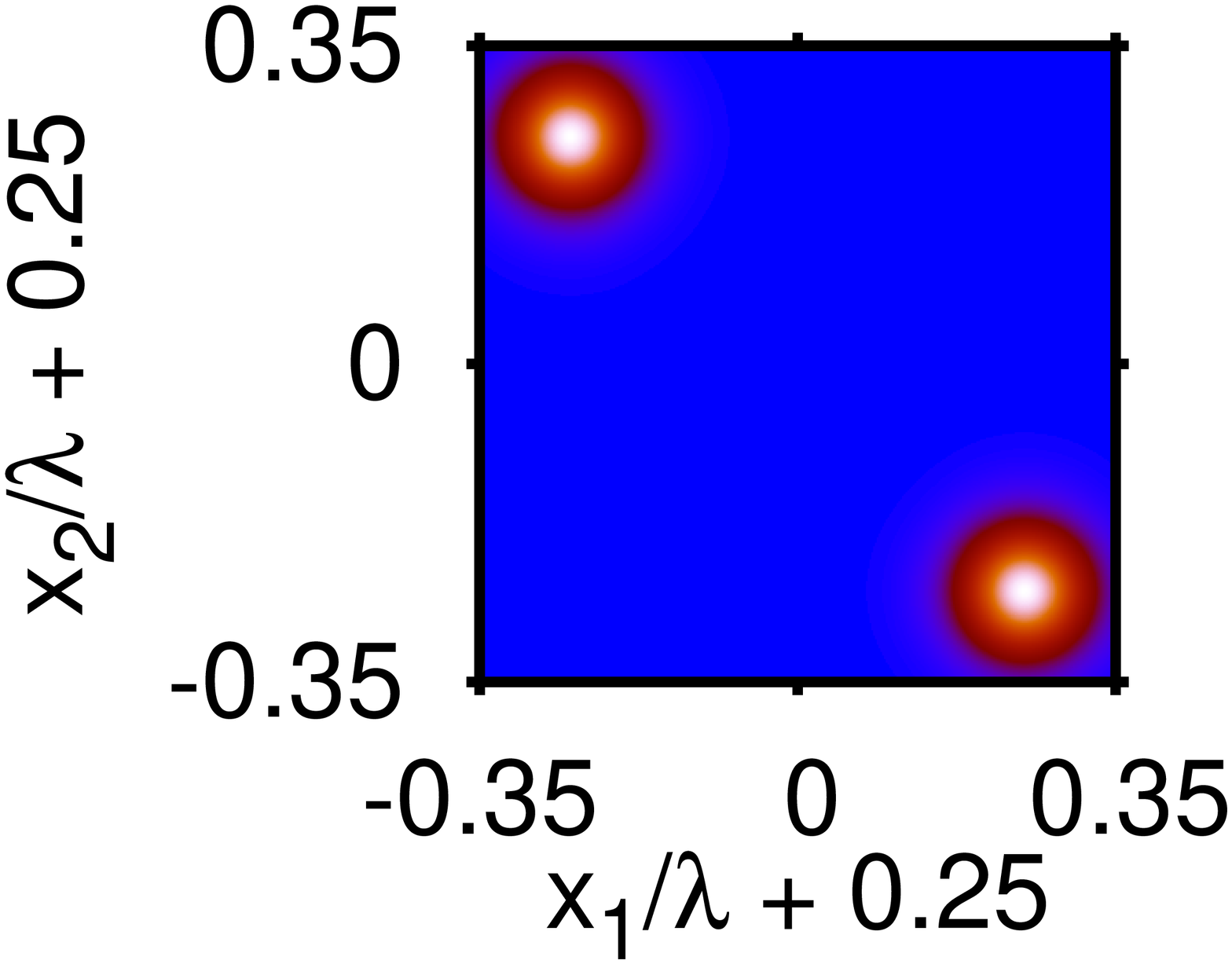}
 {\sf b}\hspace{-4mm}\includegraphics*[scale=0.17,angle=270]{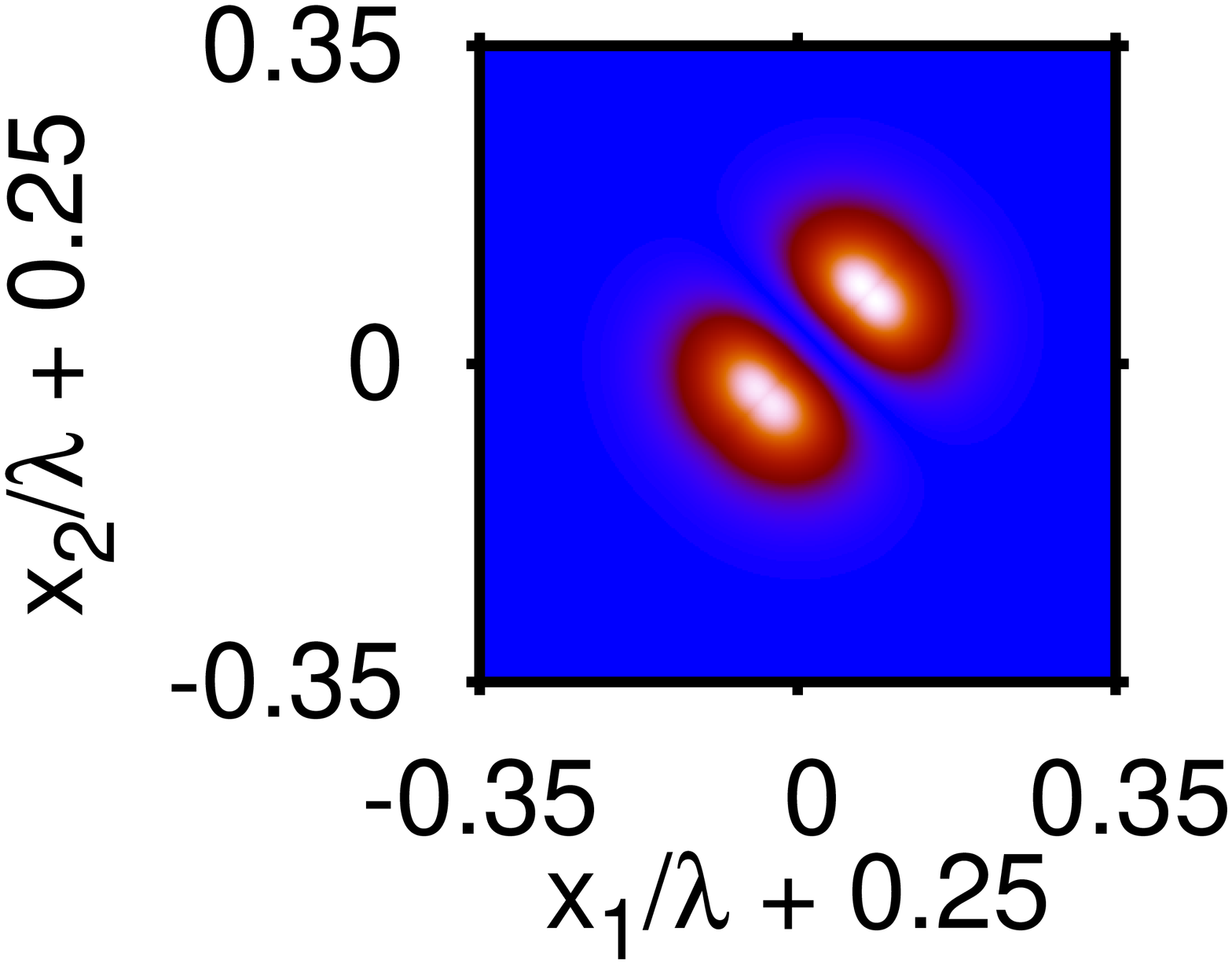}\\
 {\sf c}\hspace{-4mm}\includegraphics*[scale=0.17,angle=270]{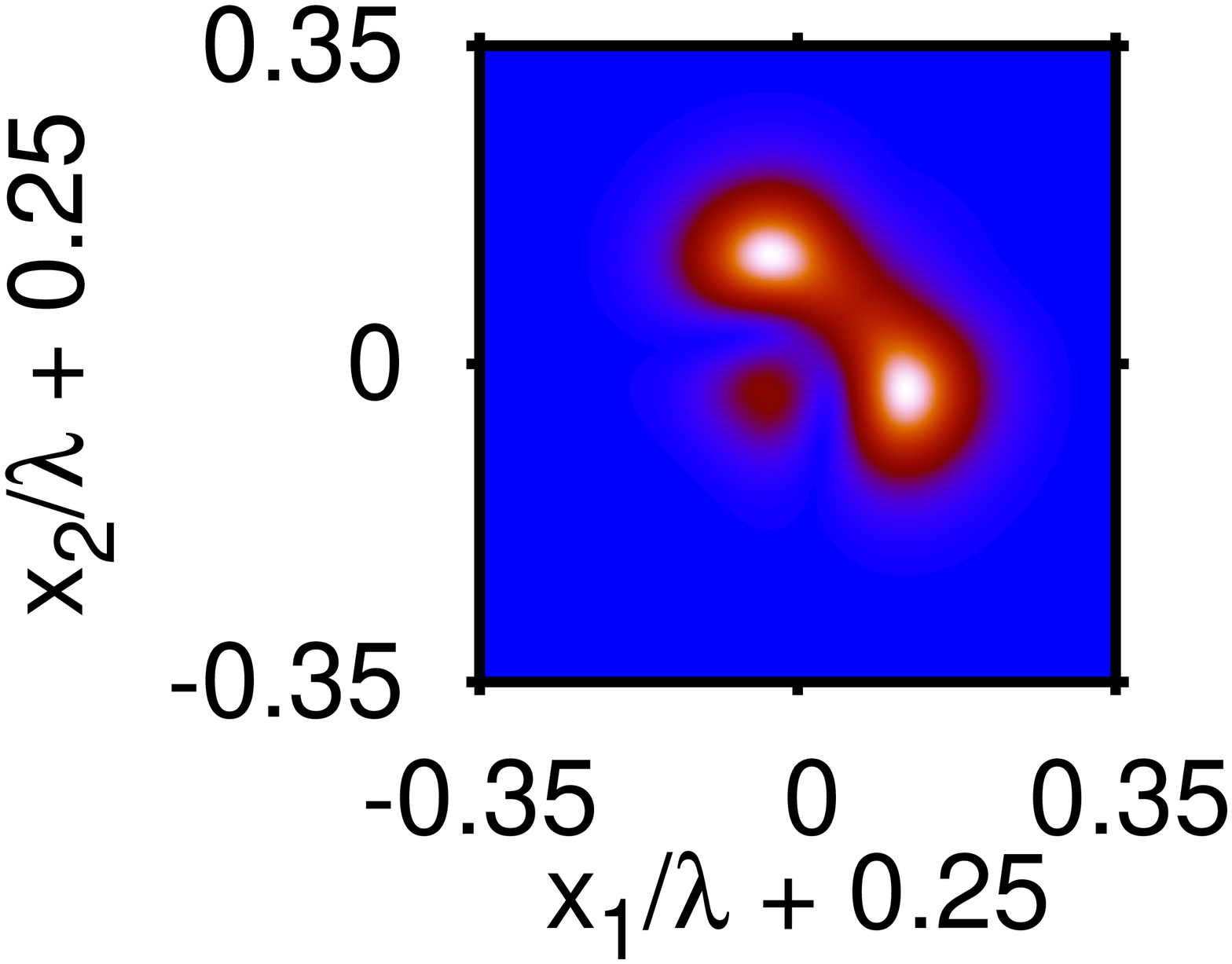}
 {\sf d}\hspace{-4mm}\includegraphics*[scale=0.17,angle=270]{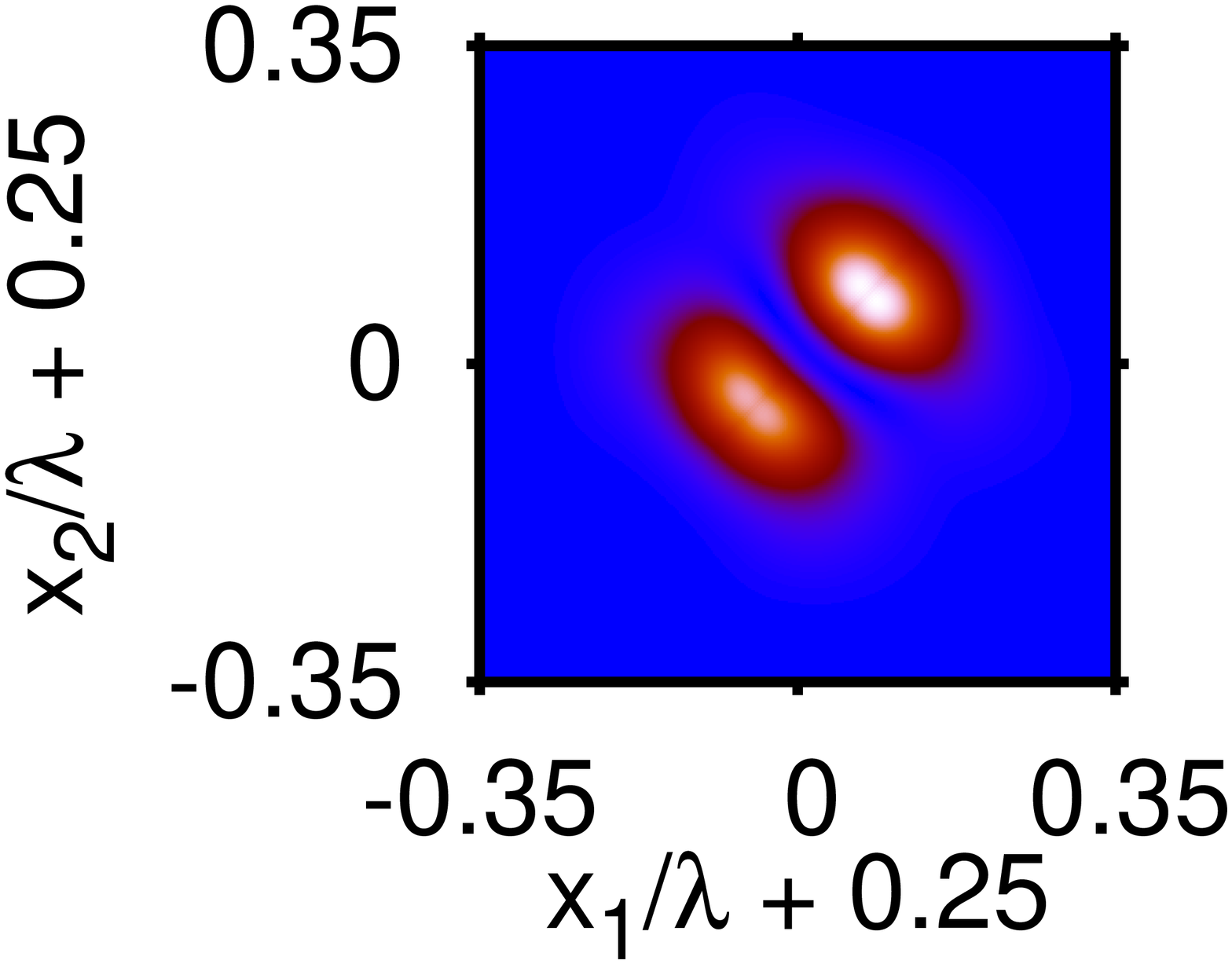}
        \caption{(Color online) Absolute square values of the relevant
        symmetric wave-functions in the coordinates of the two atoms: {\sf a}
        Initial wave function in the state $\ket{\tilde\Psi_{\rm in}}$
        with one atom in the left well and one atom in the right well; {\sf b}
        wave function of the target state $\ket{\tilde\Phi_{\rm tg}}$. {\sf c}
        evolved wave-function using the non-optimized sequences of Fig.~\ref{fig:pulse}{\sf b}
        giving a fidelity $F_{\rm int}=0.22$ (for $T=150\mu$s); {\sf d} evolved wave-function using the
        optimized sequences of Fig.~\ref{fig:pulse-opt} giving a fidelity $F_{\rm int}=0.93$.}
      \label{fig:wave-int}
\end{figure}
We consider the two-particle fidelity:
\begin{equation}
F_{\rm int}=\left|\bra{\tilde\Phi_{\mathrm{tg}}}       U_{\textrm{int}} (T)    \ket{\tilde\Psi_{\mathrm{in}}}\right|^2
\end{equation}
where $U_{\textrm{int}} (T) $ is the two-particle evolution
operator for the Hamiltonian of the two atoms, which includes
interactions. $ \ket{\tilde\Psi_{\mathrm{in}}}$ is an eigenstate of the
two-particle Hamiltonian at time $t=0$, corresponding in the limit
$g_{1D}\to 0$ to the symmetrized product of the
single-particle wavefunctions localized in each well (see
Eq.\eqref{eq:psiin}); the target state
$\ket{\tilde\Phi_{\mathrm{tg}}} $ is an eigenstate of the
two-particle Hamiltonian at time $t=T$ which, in the
limit of vanishing interactions, corresponds to the state
\begin{equation} \label{eq:psitg}
  \ket{\Phi_{\mathrm{tg}}} =\frac{1}{\sqrt 2}\left(
\ket{\phi_0}_1\ket{\phi_1}_2 + \ket{\phi_1}_1\ket{\phi_0}_2\right)
\end{equation}
The square modulus of $\ket{\tilde \Psi_{\mathrm{in}}}$ and $
\ket{\tilde\Phi_{\mathrm{tg}}}$, in the two-atom coordinate
representation, are shown in Fig.~\ref{fig:wave-int}{\sf a-b}. In order to make a comparison between the interacting and non interacting cases we define a two-particle
fidelity also in the non-interacting case:
\begin{equation}
F=\left|\bra{\Phi_{\mathrm{tg}}}       U_1 (T)\otimes U_2(T)    \ket{\Psi_{\mathrm{in}}}\right|^2
\end{equation}
where $U_1(T)$ and $U_2(T)$ are the single-particle evolution operators for the two atoms without interactions.
\begin{table}[htdp]
\begin{tabular}{|c|c|c|c|c|}
\hline
&$f_0^R$&$f_1^L$&$F$&$F_{\rm int}$\\
\hline
non optimized&0.69&0.57&0.22&0.22\\
\hline
transport optimized&0.99&0.99&0.98&0.93\\
\hline
interaction optimized&0.98&0.98&0.96&0.97\\
\hline
\end{tabular}
\caption{Results of our numerical simulations for three different sets of control parameters: the non optimized case Fig.~\ref{fig:pulse}{\sf b}; the transport optimized case Fig.~\ref{fig:pulse-opt} where the optimal control algorithm is used without taking into account interactions; the interaction optimized case where the optimal control algorithm is used taking into account interactions. The quantities shown are: the single-particle populations $f_0^R$ and $f_1^L$ calculated without interactions, the two-particle fidelities $F$ and $F_{\rm int}$ calculated without and with interactions.}
\label{tab:1}
\end{table}
In Table~\ref{tab:1} we summarize our results for $T=0.15$ ms obtained with three different sequences: first, the non optimized sequence Fig.~\ref{fig:pulse}{\sf b}; second,  the transport optimized case Fig.~\ref{fig:pulse-opt} where we used the optimal control algorithm to optimize the single-particle populations not taking into account interactions; third, the interaction optimized case where we apply the optimal control algorithm
using as the initial guess the transport optimized sequence Fig.~\ref{fig:pulse-opt} and then optimizing including the interactions in the evolution.

 The resulting
wave-functions for the non optimized and transport optimized sequences are compared in Fig.~\ref{fig:wave-int}{\sf c-d}.
Without optimal control the two-particle fidelity with and without interactions is $F=F_{\rm int}=0.22$ while with (non-interacting) optimization we obtain $F \simeq f_0^R \, f_1^L=0.98$ and $F_{\rm int}=0.93$.  This shows that interactions spoil slightly
the efficiency of the transport process as one might expect. Optimal control can subsequently be applied while including interactions in the optimization, producing a control sequence with a fidelity of
$F_{\rm int}=0.97$.

\begin{figure}[htbp]
\includegraphics[scale=0.7 ]{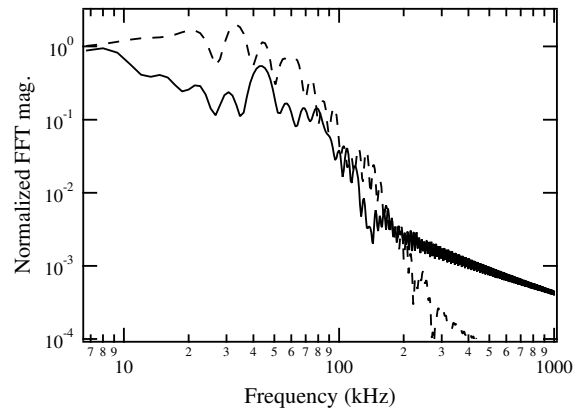}\\
         \caption{
         The normalized Fourier transform magnitudes $|\tilde{\beta}(f)|$ (solid) and $|\tilde{\theta}(f)|$ (dashed) of the optimized control sequences $\beta(t)$ and $\theta(t)$ shown in Fig.~\ref{fig:pulse-opt}. The spectra are normalized to the value at the fundamental frequency $1/T=6.67$~kHz.
                  }
      \label{fig:band}
\end{figure}

Another consideration is the experimental bandwidth available for feedback control. The optimized control waveforms Fig.~\ref{fig:pulse-opt} were obtained with no restriction on the frequency response of the control, and typically have frequency components on the order of a few times the lattice vibrational spacings (see Fig.~\ref{fig:band}), i.e. larger than the bandwidth of our control electronics. Clearly, using a filtered version of these waveforms will lead to lower control fidelity
and it will be important to increase the experimental bandwidth of the control electronics (currently about 50~kHz). In addition, it may be useful to develop an optimization sequence that includes the limited control bandwidth, although it is likely that frequencies on the order of the vibrational spacing will always be needed.

\section{Conclusions} \label{sec:conclusion}
We have presented a detailed, numerical analysis of the transport
process of neutral atoms in a time dependent optical lattice. We
show how to improve the fidelity of the transport process  for $T=0.15$ ms from
$F_{\rm int}=0.22$, using simple adiabatic switching, to $F_{\rm
int}=0.97$, using optimal control theory. We expect better results for longer control times.  We
analyze the effect of atom-atom interactions on the transport
process and we show that the optimal control parameter sequences
found in the non-interacting case still work when including
interaction. We obtained the same transformation as
in the case of the adiabatic transport with a better fidelity and
in a time shorter by more than a factor of three, which represents
a relevant improvement in terms of scalability of the number of
gates that can be performed before the system decoheres due to the
coupling to its environment. This technique can be easily adapted
to other similar transport processes and also extended to atoms in
different magnetic states, which can allow the implementation of a
fast, high-fidelity quantum gate in a real optical lattice setup
with the qubits encoded in the atomic internal states
\cite{anderliniSwap}. In the future, it would be interesting to study
the possibility of including the effect of errors in the
optimization procedure and thus investigate in more details
the robustness and noise-resilience of the optimal control
technique.

\acknowledgments This work was supported by the European
Commission, through projects SCALA (FET/QIPC), EMALI (MRTN-CT-2006-035369) and QOQIP, by
the National Science Foundation through a grant for the Institute
for Theoretical Atomic, Molecular and Optical Physics at Harvard
University and Smithsonian Astrophysical Observatory, and by DTO. SM acknowledges
support from IST-EUROSQIP and the Quantum Information program
of ``Centro De Giorgi'' of Scuola Normale Superiore.
The computations have been performed on the HPC facility of the
Department of Physics, University of Trento. We thank J.
Sebby-Strabley for experimental support.

\appendix

\section{Numerical method}
\label{sec:numerical} In our numerical simulations we employ a
finite difference method (see for example \cite{thomas}) that
consists in discretizing the coordinate representation in a
homogeneous $n$ points mesh in the interval $[X_1; X_2]$:
$x_k-x_{k-1}=dx$, $x_0=X_1$, $x_n=X_2$. The number
$dx=(X_2-X_1)/n$ is the lattice spacing. In this discretized
representation the eigenvalue equation becomes:
\begin{equation}\label{eq:eigenvalue-discrete}
 \left( V(x_k,0) -\epsilon\frac{\delta^2_{x}}{dx^2} \right)\psi_\sigma(x_k)=
 E_\sigma \psi_\sigma(x_k)
\end{equation}
where $\epsilon=3.5/(2\pi)^2 kHz$, (3.5 is the conversion
coefficient between kHz and lattice recoils) and $\psi_\sigma(x_k)$ is
the discretized wave function. The discretized second order
derivative operator $\delta^2_x$ acts on any function as:
\begin{equation}
\delta^2_x f(x_k)=f(x_{k+1})-2f(x_k)+f(x_{k-1}).
\end{equation}
Expression \eqref{eq:eigenvalue-discrete} is second order in
$dx^2$. If one arranges the function $\psi_\sigma(x_k)$ in an
$n$-dimensional array, then Eq.~\eqref{eq:eigenvalue-discrete} can
be rewritten as an eigenvalue problem with a $n\times n$
Hamiltonian matrix $H$. Since this matrix is tridiagonal, the
principal diagonal being $V(x_k,0)-2\epsilon/dx^2$ and the two
sub-diagonals being filled with $\epsilon/dx^2$, one can take
advantage of its sparse structure for storage and computing. We
calculate low energy eigenstates by approximately diagonalizing
the Hamiltonian using the Jacobi-Davidson method \cite{jacobi}.
This method is an iterative method which is capable of finding a
few (typically less than $10$) eigenstates close to a chosen
target energy. The advantages of using this and similar algorithms
(Lanczos, Arnoldi) instead of exact diagonalization are that the
method is much faster and one does not need to store the whole
Hamiltonian. In our eigenvalue problem we take full advantage of
this method, given the sparse structure of the Hamiltonian $H$.
The error of this approximate method compared to an exact
diagonalization method is negligible for our purposes.

To study the time evolution of the wave functions of the atoms we
integrate numerically the time dependent Schr\"odinger equation.
Introducing a time slicing in the interval $[0;T]$ with time
interval $dt$, the Schr\"odinger equation has the form:
\begin{equation} \label{eq:expansion-1st}
 \psi(x_k,t_{n+1})-\psi(x_k,t_{n})=-i\,dt\, H(x_k,t_n)\psi(x_k,t_{n})
\end{equation}
The discretized expression \eqref{eq:expansion-1st} gives an
iterative relation to compute the wave-function at time $t_{n+1}$,
from the expression of the wave-function at time $t_n$. This is
one example of an \emph{explicit} method: the coefficients
$\psi(x_k,t_{n+1})$ can be directly calculated from
$\psi(x_k,t_{n})$. Explicit schemes have the great advantages of
being extremely fast and easily implemented. However this
expansion is only first order in $dt$ and is not always stable.
Therefore, we used the Crank-Nicolson scheme \cite{cn}, an
\emph{implicit} method, which consists in taking a time average of
the right-hand side of \eqref{eq:expansion-1st} between time $t_n$
and $t_{n+1}$, namely
\begin{eqnarray} \label{eq:expansion-2st}
 \psi(x_k,t_{n+1})-\psi(x_k,t_{n})&=&-
 \frac{i dt}{2}\left[H(x_k,t_n)\psi(x_k,t_{n})+\right.
 \nonumber\\
 &+&
 \left . H(x_k,t_{n+1})\psi(x_k,t_{n+1})\right] \nonumber\\
 &&
\end{eqnarray}
 This method is of the second order in time and space
and it is unconditionally stable. The price for all these advantages
is that a tridiagonal set of linear equations must be solved to get
$\psi(t_{n+1})$  as shown in Eq.~\eqref{eq:expansion-2st}.  We used common Fortran routines to solve the
linear equations problem \cite{lapack}.

We solve a 2D time dependent Schr\"odinger equation in the two
coordinates of the atoms by making use of the extension in two
dimensions of the Crank-Nicolson method called the
Peaceman-Rachford method \cite{thomas}. This is an implicit method
and the integration proceeds in two steps: first the initial
wave-function is integrated in time considering only one direction
in the coordinate space, then from the intermediate wave-function
we obtain the final wave-function by integrating in the other
direction. This method is an example of alternate direction
implicit schemes.

In our simulations we used $n=(X_2-X_1)/dx=10^3$ and $n_T=T/dt=5\cdot 10^3$ that assures convergence of the results with a relative error which is less than $10^{-3}$.



\begin{thebibliography}{99}
\bibitem{bloch-review}
For a review, see for example: I. Bloch, J. Phys. B {\bf 38}, S629  (2005).
\bibitem{inguscio-fermi}
G. Modugno, F. Ferlaino, R. Heidemann, G. Roati, M. Inguscio,
Phys. Rev. A {\bf 68}, 011601(R) (2003).
\bibitem{bloch-qpt} M. Greiner, O. Mandel, T. Esslinger, T. W. H\"{a}nsch,
and I. Bloch, Nature London  {\bf 415}, 39 (2002).

\bibitem{cirac-chains}
J. J. Garc\`ia-Ripoll, M. A. Martin-Delgado, and J. I. Cirac
Phys. Rev. Lett. 93, 250405 (2004).
\bibitem{brennen}
G. K. Brennen, C. M. Caves, P. S. Jessen, and I. H. Deutsch,
Phys. Rev. Lett. {\bf 82}, 1060  (1999).
\bibitem{jaksch}
 D. Jaksch, H.-J. Briegel, J. I. Cirac, C. W. Gardiner, and P.
Zoller, Phys. Rev. Lett. {\bf 82}, 1975  (1999).
\bibitem{dorner} T. Calarco, U. Dorner, P. Julienne, C. Williams,
and P. Zoller, Phys. Rev. A {\bf 70}, 012306  (2004).
\bibitem{lee07}
P. J. Lee, M. Anderlini, B. L. Brown, J. Sebby-Strabley, W. D.
Phillips and J. V. Porto, Phys. Rev. Lett. {\bf 99}, 020402 (2007).

\bibitem{mandelNature03}
O. Mandel, M. Greiner, A. Widera, T. Rom, T. W. H\"{a}nsch and I.
Bloch, Nature \textbf{425}, 937 (2003).
\bibitem{oc1}
A. P. Peirce, M. A. Dahleh, and H. Rabitz, Phys. Rev. A {\bf 37},
4950 (1988).  R. Kosloff, S. A. Rice, P. Gaspard, S. Tersigni, and
D. J. Tannor, Chem. Phys. {\bf 139}, 201 (1989).
\bibitem{trotzky08a}
S. Trotzky, P. Cheinet, S.Folling, M. Feld, U. Schnorrberger, A. M. Rey, A. Polkovnikov, E. A. Demler, M. D. Lukin, and I. Bloch, Science {\bf 319}, 295 (2008).

\bibitem{anderliniSwap} M. Anderlini, P. J. Lee, B. L. Brown, J. Sebby-Strabley,
W. D. Phillips and J. V. Porto, Nature {\bf 448} 452--456 (2007).

\bibitem{garcia-2003}
J. J. Garc\`ia-Ripoll, P. Zoller, and J. I. Cirac,
Phys. Rev. Lett. {\bf 91}, 157901 (2003).
\bibitem{anderlini-pra}
J. Sebby-Strabley, M. Anderlini, P. S. Jessen, and J. V. Porto,
Phys. Rev. A {\bf 73}, 033605 (2006).
\bibitem{anderliniJPhysB} M. Anderlini, J. Sebby-Strabley, J. Kruse, J.V.
Porto, and W.D. Phillips, J. Phys. B  \textbf{39}, S199 (2006).
\bibitem{SebbyPRL07} J. Sebby-Strabley, B. L. Brown, M. Anderlini, P. J. Lee,
W.D. Phillips, J. V. Porto, and P. R. Johnson, Phys. Rev. Lett.
{\bf 98}, 200405 (2007).

\bibitem{kastberg95} A. Kastberg, W. D. Phillips, S. L. Rolston, R. J. C. Spreeuw, and
P. S. Jessen, Phys. Rev. Lett. \textbf{74}, 1542 (1995).
\bibitem{greiner2001}
M. Greiner, I. Bloch, O. Mandel, T. W. H\"ansch, and T. Esslinger,
Phys. Rev. Lett. \textbf{87}, 160405 (2001).
\bibitem{Ovchinnikov1998}
Yu. B. Ovchinnikov, J. H. M\"uller, M. R. Doery,
 E. J. D. Vredenbregt, K. Helmerson, S. L. Rolston, and W. D. Phillips,
  Phys. Rev. Lett \textbf{83}, 284 (1999).
\bibitem{spielman06}
I. B. Spielman, P. R. Johnson, J. H. Huckans, C. D. Fertig, S. L.
Rolston, W. D. Phillips, and J. V. Porto Phys. Rev. A {\bf 73},
020702(R) (2006).
\bibitem{thomas}
J. W. Thomas: ``Numerical Partial Differential Equations'', vol.
1, Springer, New York (1995).
\bibitem{cn}
J. Crank and P. Nicolson, Proc. Cambridge Philos. Soc. {\bf 43}, 50 (1947).



\bibitem{oc2}
S. A. Rice and M. Zhao: ``Optical control of Molecular dynamics'', J. Wiley, New York (2000).
\bibitem{oc3}
M. Shapiro and P. Brumer: ``Principles of the quantum control of
molecular processes'', J. Wiley, New York (2003).
\bibitem{tannor-2002}
S. E. Sklarz and D. J. Tannor, Phys. Rev. A {\bf 66}, 053619 (2002).
\bibitem{vaucher}
B. Vaucher, S. R. Clark, U. Dorner, D. Jaksch, New J. Phys. {\bf 9} 221 (2007).
\bibitem{romero}
O. Romero-Isart and J. J. Garc\`ia-Ripoll, Phys. Rev. A {\bf 76}, 052304 (2007).
\bibitem{monta-oc}
S. Montangero, T. Calarco and R. Fazio,
Phys. Rev. Lett. {\bf 99}, 170501 (2007).
\bibitem{krotov}
V. F. Krotov: ``Global methods in optimal control theory'', M. Dekker Inc., New York (1996).

\bibitem{olshanii}
M. Olshanii, Phys. Rev. Lett. {\bf 81}, 938  (1998).


\bibitem{jacobi}
G. L. G. Sleijpen and H. A. van der Vorst, SIAM Review {\bf 42},
267 (2000), we used the routine written by G. L. G. Sleijpen and
coworkers: http://www.math.uu.nl/people/sleijpen/
\bibitem{lapack}
E. Anderson et al.: ``LAPACK User's guide'', SIAM, Philadelphia
(1999).
\end{thebibliography}
\end{document}